\documentstyle[12pt,aasms4]{article}

\slugcomment{Submitted to the Astrophysical Journal, 6 March 1997}

\lefthead{Isenberg, Lamb, and Wang} \righthead{Effects of the Geometry
of the Line-Forming Region}

\begin{document}

\title{Effects of the Geometry of the Line-Forming Region on the
Properties of Cyclotron Resonant Scattering Lines}

\author{Michael Isenberg and D. Q. Lamb} \affil{Department of Astronomy
and Astrophysics, University of Chicago, 5640 South Ellis Avenue,
Chicago, IL 60637}

\author{John C. L. Wang} \affil{Department of Astronomy, University of
Maryland, College Park, MD 20742-2421}

\begin{abstract}
We use a Monte Carlo radiative transfer code to examine the dependence
of the properties of cyclotron resonant scattering lines on the spatial
geometry and the optical depth of the line-forming region.  We focus
most of our attention on a line-forming region that is a plane-parallel
slab.  We also consider a cylindrical line-forming region.  In both
cases, the line-forming region contains an electron-proton plasma at
the equilibrium Compton temperature, $T_c$, and is threaded with a
uniform magnetic field with strength $\sim 10^{12}$ gauss.  We consider
geometries in which the photon source illuminates the line-forming
region from below, and in which the photon source is embedded in the
line-forming region.  The former may correspond to a line-forming region
in the magnetosphere of a neutron star, illuminated from below; the
latter to a line-forming region on or near the surface of a neutron
star as in an accretion column.  

In the case of the plane-parallel slab line-forming region, we study
the behavior of the resonant Compton temperature and the properties of
the cyclotron scattering lines as a function of the column depth of the
line-forming region and the orientation of the magnetic field.  At
small or moderate optical depths ($N_e \ll 10^{25}\ {\rm cm^{-2}}$
[electron scattering optical depth $\tau_{To} \ll 10$]), the resonant
Compton temperature ranges from $T_e/E_B \approx 0.2$ when the magnetic
field is perpendicular to the slab to $T_e/E_B \approx 0.4$ when the
field is parallel to the slab.  At large optical depths ($N_{e} \gtrsim
1.5\times10^{25} {\rm cm^{-2}}$ [$\tau_{To} \gtrsim 10$]), the resonant
Compton temperature is higher.

At small or moderate optical depths ($\tau_{To} \ll 10$), the
equivalent widths, $W_E$, of the spectral lines depend on two effects.
First, $W_E$ increases as the viewing angle {\it with respect to the
field} decreases, due to Doppler broadening.  Second, $W_E$ increases
as the viewing angle {\it with respect to the slab normal} increases,
due to the increase in column depth along the line of sight.  When the
magnetic field is parallel to the slab and the viewing angle is along
the field, for example, both effects are large; in a line-forming
region whose optical depth is moderate, the first harmonic equivalent
width can reach $\sim 0.75E_B$.  Shoulders are present on each side of
the first harmonic line when the spectrum is viewed at angles with
respect to the slab normal corresponding to $\mu_{sl}=\cos \theta_{sl}
\gtrsim 0.25$ ($\theta_{sl}\lesssim 75^{\rm o}$), and become stronger
with increasing column depth, until column depths $N_{e} \lesssim
1.5\times10^{25} {\rm cm^{-2}}$.  As a result, the $W_E$ of the first
harmonic reaches a maximum value, and then decreases with increasing
column depth (optical depth) in the line-forming region.  The shoulders
are the result of injecting the photon spectrum at moderate optical
depths.  When the photon spectrum is injected at sufficiently large
optical depths ($\tau_{To} \gtrsim 10$), continuum scattering smears
the shoulders so that they are no longer visible and the equivalent
width of the first harmonic resumes increasing with increasing column
depth.

In the case of the cylindrical line-forming region, we consider only a
magnetic field oriented along the axis of the cylinder.  This
corresponds to the situation expected in the canonical model of the
emission region of accretion-powered pulsars.  We find that the
emerging spectrum is similar to that of the slab line-forming region
with an embedded photon source and the magnetic field perpendicular to
the slab normal.  Thus, at large optical depths we expect the cyclotron
line from a cylindrical line-forming region to have a very large
equivalent width $W_E$, nearly independent of viewing angle, and no
visible shoulders.

Our findings have implications for accretion-powered pulsars and
gamma-ray bursters.  In particular, the absence of pronounced shoulders
on each side of the cyclotron first harmonic line in the spectra of
accretion-powered pulsars suggests that the line-forming region is either
illuminated from below and outside, as would be the case if it were
plasma suspended in the magnetosphere of the neutron star, or it has a
large ($\tau_{To} \gtrsim 10$) optical depth.  Also, the ability of a
slab line-forming region in which the magnetic field is parallel to the
slab to produce narrow lines with large $W_E$ suggests that the lines
observed in the X-ray spectra of some gamma-ray bursts might be able to
be formed in plasma trapped at the magnetic equator of a neutron star.
\end{abstract}

\section{Introduction}
\label{intro}

\medskip
Cyclotron lines in an astrophysical x-ray spectrum are the clear
signature of a superstrong magnetic field in the line-forming region.
Consequently, they are an important clue to the nature of the source
and the properties of the emission region, especially at energies
$\gtrsim\ 10$ keV where atomic lines are not available.  Cyclotron
lines have been important in identifying accreting magnetized neutron
stars as the source of radiation in  about a dozen X-ray pulsars
(Mihara 1995; Makishima and Mihara 1992), including Her X-1 (Tr\"umper
{\it et al.} 1978), 4U 0115+63 (Wheaton {\it et al.} 1979), 4U 1538-52
(Clark {\it et al.} 1990), and A0535+26 (Grove {\it et al.} 1995).
Absorption-like features have also been observed in classical gamma-ray
bursts and interpreted as cyclotron lines (Mazets {\it et al.} 1980,
1981; Heuter 1984; Murakami {\it et al.} 1988; Fenimore {\it et al.}
1988; Harding and Preece 1989; Wang {\it et al.} 1989; Alexander and
M\'esz\'aros 1989; Lamb {\it et al.} 1989; Yoshida {\it et al.} 1991;
Nishimura and Ebisuzaki 1992; Nishimura 1994; Mazets {\it et al.}
1996; Briggs 1996).  Although the
observation of spectral features in gamma-ray bursts and their
interpretation continue to be controversial, they are perhaps the
strongest evidence that at least some gamma-ray bursts come from
strongly magnetized neutron stars associated with our own galaxy.

\medskip
By building theoretical models of line-forming regions and comparing
the emerging spectra with observations, theorists have been able to
infer {\it quantitatively} many properties of the sources that are
believed to possess cyclotron lines.  For example, M\'esz\'aros and
Nagel (1985) use a Feautrier calculation to show that the spectrum of
Her X-1 could be produced in a region with magnetic field $B_{12}
\equiv B/(10^{12} {\rm gauss})$=3.3 oriented parallel to the normal of
a slab atmosphere with column depth (slab width) $N_{e,21} \equiv
N_e/(10^{21}\ {\rm electrons\ cm^{-2}})=10^4$ and electron temperature
$T_e$=7 keV.  Other fits to accretion-powered pulsar spectra include
those of \cite{ah96}, who infer the presence of a magnetic field
approaching the critical field, $B_{crit,12}=44.14$ from the spectrum
of A0535+26.  For gamma-ray bursts, \cite{wang89} use a Monte Carlo
radiative transfer code (Wang, Wasserman, and Salpeter 1988; Lamb {\it
et al.} 1989) to show that the cyclotron lines in the spectrum of the
burst GB880205 could be produced in a region with $B_{12}$=1.7,
$N_{e,21}$=1.2, and with electrons at the equilibrium Compton
temperature, $T_c$, which is 5.3 keV in this case.

\medskip
\cite{wang89} inject an initial photon distribution into a
plane-parallel slab atmosphere from a source plane below the slab.
Here we consider both this model and one in which the photons are
injected at a source plane embedded inside the slab.  A slab
illuminated from below may correspond to a line-forming region in the
magnetosphere of a neutron star (cf. Dermer and Sturner 1991; Sturner
and Dermer 1994; see also Zheleznyakov and Serber 1994, 1995).  A slab
illuminated from within may correspond to a line forming region in a
semi-infinite atmosphere at the stellar surface (see Slater, Salpeter,
and Wasserman 1982; Wang, Wasserman, Salpeter 1988, 1989; Freeman {\it
et al.} 1992; Nelson {\it et al.} 1997).

\medskip
\cite{wang89} take the magnetic field to be oriented along the slab
normal, $\hat z_{sl}$ (i.e. $\Psi=0$).  Such fields are found near the
magnetic poles of neutron stars with dipole fields.  Kaminker, Pavlov
and Shibanov (1982, 1983) and \cite{bak91} consider the effects of
varying the angle $\Psi$ between the magnetic field and the slab
normal.  However, these authors do not include the effects of
scattering in the cyclotron line core (see Section \ref{secmc});
\cite{bak91} calculate the spectrum in the continuum only, while
Kaminker, Pavlov, and Shibanov's (1982, 1983) calculation is valid for
the continuum and the line wings.

\medskip
In the present work, we consider scattering in line-forming regions
with $\Psi \neq 0$, {\it including} core scattering.  Fields at angles
other than $\Psi=0$ occur when lines are formed in structures near the
magnetic poles, such as accretion columns or mounds (Burnard, Arons,
and Klein 1991).  Line-forming regions with $\Psi \neq 0$ also
correspond to non-polar regions on the surface or in the magnetosphere
of neutron stars with dipole fields, as well as regions on stars with
more complicated fields.  For example, the surface of a neutron star
may be threaded by local fields with various orientations relative to
the local slab normal, instead of or in addition to a global dipole
field (see, e.g., Ruderman 1991a,b,c; Lamb, Miller, and Taam 1996).

\medskip
We will show that if the magnetic field is oriented perpendicular
to the slab normal ($\Psi = \pi/2$) and the photon source is embedded
in the slab, the line properties are similar to those for a
cylindrical line-forming region.  Thus a slab line-forming region with
this geometry is a good approximation to the cylindrical line-forming
region expected in the canonical model of accretion-powered pulsars.

\medskip
In Section \ref{seclfr} we discuss the geometry of the line-forming
regions we use in our calculations.  In Section \ref{secmc} we
describe the Monte Carlo radiative transfer code.  We show how the
equilibrium Compton temperature varies with the field orientation in
Section \ref{secCompton}.  In Section \ref{secSpectra} we present the
spectra calculated by the Monte Carlo code and discuss the effects of
the line-forming region geometry on the properties of the emergent
lines.  Finally, in Section \ref{secDisc} we discuss the implications
of our calculations for accretion-powered pulsars and gamma-ray
bursters.

\section{Spatial Geometry of the Line-Forming Region}
\label{seclfr}

\subsection{Slab Geometry}

The principal spatial geometry we study in this paper, a plane parallel
slab, is shown in Figure \ref{geometry}a.  Photons are injected at the
source plane, travel through an electron-proton plasma, and emerge from
one or the other of the faces of the slab.  The horizontal extent of
the slab is infinite.  The angle $\theta$ is the polar angle of a
photon's direction of propagation {\it with respect to the magnetic
field}.  The polar angle is usually specified in terms of $\mu \equiv
\cos\theta$.  Similarly, $\theta_{sl}$ and $\varphi_{sl}$ are the polar
and azimuthal angles, respectively, of a photon {\it with respect to
the slab normal} and $\mu_{sl}\equiv\cos\theta_{sl}$.

\medskip
As the Figure shows, $N_e$ is the column depth between the top of the
slab and the source plane, while $N_e^\prime$ is the column depth
between the source plane and the bottom of the slab.  In some
circumstances, we will also specify the column depth in terms of the
Thomson depth, $\tau_{To} \equiv N_e \sigma_T $, where
$\sigma_T \equiv\ (8\pi/3)(e^2/(m_e c^2))^2 = 6.65\times10^{-25}\ {\rm
cm^2}$ is the Thomson cross section.  We stress that in general the
optical depth encountered by a given photon will differ from the
Thomson depth since the actual scattering cross section varies with
photon energy, direction, and polarization.

We characterize the position of the source plane by the ratio of $N_e$
to $N_e^\prime$.  Thus, for a slab illuminated from below, as in Wang
{\it et al.} (1989), the column depth between the source plane and the
bottom of the slab is zero; we call this the 1-0 geometry.  Similarly,
a slab with the source plane embedded in the middle of the slab, so
that $N_e\ =\ N_e^\prime$, has a 1-1 geometry.

Slater, Salpeter, and Wasserman (1982) and Wang, Wasserman, and
Salpeter (1988) determine that the mean number of scatters between the
source and the top edge that a resonant photon experiences prior to
escape approaches a limiting value as the atmosphere becomes
semi-infinite (i.e. as $N_e^\prime/N_e \to \infty$).  It is generally
within ten percent of its limiting value in the 1-1 geometry and
indistinguishable from its limiting value in the 1-4 geometry.  The
corresponding mean path length the photon travels between the source
and the top edge displays similar behavior, as does the median
frequency shift of the escaping photon.  Thus the 1-4 geometry is an
excellent approximation to a semi-infinite atmosphere.  Unfortunately,
our Monte Carlo simulations show that the total number of scatters per
injected photon in the 1-4 geometry is approximately three times the
number of scatters in the 1-1.  The computer time required for the
simulation increases proportionately.  Consequently, we use the 1-1
geometry in the present work as a compromise between the quality of the
approximation to a semi-infinite atmosphere and the amount of computer
time required.

We refer to photons that emerge from the top of the slab as {\it
transmitted} and those that emerge from the bottom as {\it reflected}.
The former reach the observer.  In the 1-0 geometry, we assume that the
latter return to the stellar surface where they are thermalized, i.e.,
absorbed by nonresonant inverse magnetic bremsstrahlung.  In the 1-1
geometry the reflection symmetry of the line-forming region across the
source plane ensures that the transmitted and reflected spectra for
isotropically injected photons are the same.
We can, therefore, set the transmitted spectrum for isotropic injection
equal to the {\it sum\/} of the transmitted and reflected spectra for
{\it semi}-isotropic injection.  By doing so, we can reduce the computer
time required per transmitted photon by an additional factor $\sim 2$
in the 1-1 geometry as compared with the 1-4 geometry.

\subsection{Cylindrical Geometry}

We also consider a cylindrical geometry, as shown in Figure
\ref{geometry}b.  The length of the cylinder is infinite.  We only
consider magnetic fields oriented parallel to the cylinder axis, as
expected in the canonical model of the emission region of
accretion-powered pulsars.  Photons are injected along the cylinder
axis and emerge from the surface of the cylinder.  As in the slab
geometry, the angle $\theta$ is the polar angle of a photon's
direction of propagation with respect to the magnetic field.  The
column depth between the cylinder axis and the surface of the cylinder
is $N_e$.

\section{Monte Carlo Radiative Transfer Code}
\label{secmc}

\medskip
Our Monte Carlo radiative transfer code is an extended version of that
used by Wang {\it et al.}  (1989).  The code injects an initial photon
distribution $N_i(E)$ into an isothermal, fully ionized slab atmosphere
with a uniform magnetic field at an angle $\Psi$ to the slab normal.
It then follows each photon through the slab and determines the
emergent spectrum.

\medskip
The code uses resonant scattering cross sections that have been
averaged over the initial polarization states and summed over the final
states.  \cite{wws88} argue that polarization averaged cross sections
are appropriate for first harmonic scattering in optically thick media
when the vacuum contribution to the dielectric tensor dominates the
plasma contribution.  This will be the case when

\begin{equation}
{w \over \delta}=0.04 \left ({n_e \over 10^{20}\ {\rm cm^{-3}}}\right) B_{12}^{-4} \ll 1,
\end{equation}

\noindent
where $w \equiv (\hbar \omega_p/E_B)^2$ is the plasma frequency
parameter, $\omega_p$ is the plasma frequency, $E_B=11.6\ {\rm keV}\
B_{12}$ is the cyclotron energy, $\delta$ is the magnetic vacuum
polarization parameter (see, e.g., Adler 1971), and $n_e$ is the
electron number density.  This condition will generally hold under the
physical conditions studied in this paper.

\medskip
We work in the cold plasma approximation,
 in which $kT_e \ll E_B$. 
In this approximation, the Landau level spacing is much larger than the
typical electron energy so the Landau levels are not collisionally populated. 
We further assume that the photon densities are sufficiently low so that the 
levels are not radiatively populated. Thus, in each scattering the
initial and final 
electron state is the Landau ground state ($n=0$).  
We denote scattering
channels by the sequence of Landau levels that the electron occupies,
e.g. $0 \rightarrow 0$, $0\rightarrow 1\rightarrow 0$,
 and $0 \rightarrow 3 \rightarrow 2 \rightarrow 0$.

In our treatment of electron-photon scattering, we adopt the approximation

\begin{equation}
\left\vert \sum_i a_i \right\vert^2 \approx \left\vert a_{0
\rightarrow 0} + a_{0 \rightarrow 1 \rightarrow 0} \right\vert^2 +
\sum_{i\neq 0 \rightarrow 0, \atop 0 \rightarrow 1 \rightarrow 0} \left\vert a_i
\right\vert^2\ ,
\label{BasicAssump}
\end{equation}

\noindent
where $a_i$ is the matrix element for the $i^{th}$ scattering channel.
In general, this approximation is a good one only near the line
centers.  Consequently, eq. (2) is valid for line-forming regions which
are not optically thick in the line wings of the first harmonic (see
below; also Wasserman \& Salpeter 1980; Lamb {\it et al.} 1989), that
is

\begin{equation}
\tau_1 \lesssim 1/a\ ,
\end{equation}

\noindent
where 

\begin{equation}
\tau_1 = 100 N_{e,21}B_{12}^{-1}\left({T_e \over 1\ {\rm keV}}\right)^{-1/2}
\label{tau1}
\end{equation}

\noindent
is the polarization-, angle-, and frequency-averaged optical depth in
the first harmonic,

\begin{equation}
a \equiv {\Gamma_1 \over 2 E_d^1}=1.8 \times 10^{-3}\ B_{12} \left({T_e \over
1\ {\rm keV}}\right )^{-1/2}
\end{equation} 

\noindent
is the dimensionless natural line width,

\begin{equation}
E_d^n \equiv n E_B \sqrt{2 T_e \over m_e c^2} = 0.73\ n B_{12}  \left({T_e \over
1\ {\rm keV}}\right )^{1/2}
\end{equation}

\noindent 
is the Doppler width associated with the $n^{th}$ harmonic, 

\begin{equation} 
\Gamma_n = {4 n \alpha E_B^2 \over 3 m_e c^2}=2.6 \times 10^{-3} \ {\rm keV}\ nB_{12}^2
\end{equation}

\noindent
is the radiative width for the $n^{th}$ harmonic, and $\alpha$ is the
fine structure constant.  Even when the line-forming region is
optically thick in the line wings, eq. (\ref{BasicAssump}) is still a
good approximation if the spectrum has an exponential rollover such
that there are very few photons beyond the first harmonic.  This is
usually the case for the spectra of accretion-powered pulsars (Mihara
1995).  Thus, for instance, for rollover energies of $kT_\gamma
\approx E_B/4$, the number of photons with energies significantly
above the first harmonic is negligible.  In this case only continuum
and first harmonic scattering will be significant.  However, if
neither of these conditions hold, we cannot ignore the cross terms
when squaring the matrix element, as we have in
eq. (\ref{BasicAssump}).  For the most part, in the present work, we
confine ourselves to cases where eq. (\ref{BasicAssump}) is valid.
However, we shall see below that when the magnetic field is parallel
to the slab, photons traveling along the slab see a very large column
depth {\it and\/} a large cross section and so can scatter multiple
times with a wide range of electron velocities.  Some of these photons
can therefore continue to scatter far in the line wings above $\sim
E_B$ where eq. (\ref{BasicAssump}) no longer holds.
However, the range of magnetic field orientations and photon
directions where this occurs is small.

\medskip
The scattering cross section we use is valid in the limit $(E^r / m_e
c^2)^2 b^{-1} \ll 1$, where $b \equiv B/B_{crit}$, $E$ is the photon
energy, and the superscript $r$ denotes quantities measured in the
frame of reference where the electron is at rest ($p_z = 0$) prior to
scattering, that is, the pre-scattered electron's rest frame (see,
e.g., Daugherty and Ventura 1977).  This limit applies throughout the
present work.  For the $0 \rightarrow 0$ and $0 \rightarrow 1
\rightarrow 0$ channels, these limits give the classical magnetic
Compton scattering cross section.  When evaluated in the pre-scattered
electron's rest frame and averaged over the azimuthal angle $\varphi$,
this cross section is given by (Canuto, Lodenquai, Ruderman 1971;
Herold 1979; Ventura 1979; Wasserman and Salpeter 1980; Harding and
Daugherty 1991; Graziani 1993; first term on right-hand-side of
eq. [\ref{BasicAssump}]):

\begin{equation}
{d\sigma^r \over d\Omega^r_s} = {d\sigma^r_{0 \rightarrow 0} \over d\Omega^r_s}
+ {d\sigma^r_{0 \rightarrow 1 \rightarrow 0} \over d\Omega^r_s},
\label{ClassCross}
\end{equation}

\noindent
where 
\begin{equation}
{d\sigma^r_{0 \rightarrow 0} \over d\Omega^r_s} \equiv {3 \over 16
\pi} \sigma_T \left[ \sin^2 \theta^r \sin^2 \theta^r_s + \left( {E^r \over
E^r + E_B} \right)^2 \left({1+\mu^{r2} \over 2}\right)
\left(1+\mu^{r2}_s\right) \right]
\label{NonResCross}
\end{equation}

\noindent
is the continuum part of the cross section and

\begin{equation}
{d\sigma^r_{0 \rightarrow 1 \rightarrow 0} \over d\Omega^r_s} \equiv
{9 \over 32} {\sigma_T m_e c^2 \over \alpha} 
\left [   {4 E^{r3} \over E_B (E^r+E_B)^2} \right ]
{\Gamma_1/2\pi \over (E^r-E^r_1)^2 + (\Gamma_1/2)^2} {1+\mu^{r2} \over 2}
{1+\mu^{r2}_s \over 2}
\label{ResCross}
\end{equation}

\noindent
is the resonant part.
We
use the exact relativistic resonant energy for the $n^{th}$ harmonic $E_n^r=
  2nb m_e c^2/[1+\sqrt{1+2nb(1-{\mu^r}^2)}]$ with $n=1$ in the
Lorentzian factor in eq. (\ref{ResCross}).  
The subscript $s$ denotes parameters of the scattered photon.

\medskip
For a $0 \rightarrow 0$ scattering we obtain the total continuum scattering
cross section $\sigma_0^r$ by integrating eq. (\ref{NonResCross}) 
over the scattered angle.  We define $\sigma_n^r$ as the total cross
section for the resonant absorption that initiates a (resonant)
scattering at the $n^{th}$ harmonic.  For $n\geq 1$, this is given by
(Daugherty and Ventura 1977; Fenimore {\it et al.} 1988):

\begin{equation}
\sigma_n^r = {3 \over 8}
{\pi m_e c^2 \sigma_T \over \alpha} b^{n-1} {(n^2/2)^{n-1} \over
(n-1)!} (1 + \mu^{r2} 
) (1 - \mu^{r2})^{n-1} {\Gamma_n/2\pi \over (E^r-E^r_n)^2 + (\Gamma_n/2)^2}.
\label{sigrest}
\end{equation} 

\noindent
In order to achieve greater accuracy in the first harmonic line wings,
we use the value of $\sigma^r_1$ obtained by integrating eq.
(\ref{ResCross}) over the scattered angle rather than using eq.
(\ref{sigrest}) which is strictly valid only near line center.  The two
expressions converge when $E^r \to E^r_1$.  For the higher harmonics,
we use eq. (\ref{sigrest}).

\medskip
 From the Lorentz invariance of $\tau/\vert\mu_{sl}\vert$ (see, e.g.,
 Rybicki and Lightman 1979), the lab frame cross section $\sigma_n$ is
related to $\sigma_n^r$ by

\begin{equation}
\sigma_n = (1-\beta\mu)\,\sigma_n^r\ .
\label{siglab}
\end{equation} 

\medskip\noindent
We average this cross section over $f(p)$, the one-dimensional
electron momentum distribution, and divide by $\sigma_T$ in order to
obtain the dimensionless scattering profile for the $n^{th}$ harmonic,

\begin{equation}
\phi_n(E,\Omega) \equiv {1 \over \sigma_T}\ \int dp f(p) \sigma_n\ .
\label{NScatProfile}
\end{equation}

\medskip\noindent
The total scattering profile is approximately the sum of the profiles for the 
individual Landau levels (including the continuum contribution):

\begin{equation}
\phi(E,\Omega) \equiv {1 \over \sigma_T} \int{ dp f(p) \int d\Omega_s
{d\sigma \over d\Omega_s }} \approx \sum_{n=0}^\infty
\phi_n (E,\Omega)
\label{ScatProfile}
\end{equation}

\noindent
(Wasserman and Salpeter 1980; Wang, Wasserman, and Lamb 1993).

\medskip
Figures \ref{profile}a, b, and c show the scattering profiles
$\phi_0(E,\Omega)$, $\phi_1(E,\Omega)$, and
$\phi_0(E,\Omega)+\phi_1(E,\Omega)$ vs. $E$ for $B_{12}=3.5$, $T_e=10$
keV, and $\mu=$0, 0.5, and 1, respectively.  It is clear from the
figure that the line can be divided into the line core ($|x_n/\mu| \ll
1$) and the line wings ($|x_n/\mu| \gg 1$), where $x_n \equiv (E -
E_n)/E_d^n$ is the dimensionless frequency shift (Wasserman and
Salpeter 1980).   In the line core, the thermal electron distribution
dominates the profile so that $\phi_n \propto \exp[-(x_n)^2/\vert \mu
\vert]$.  In the wings, the tail of the Lorentzian distribution
dominates so that $\phi_n \propto  a (x_n)^{-2}$.  We refer to the wing
at energies below the line center as the red wing and the wing at
energies above the line center as the blue wing.  \cite{ws80} showed
that for the first harmonic, the core-wing boundary appears at
$|x_1/\mu| \approx 2.62 - 0.19 \ln(100\ a/\mu)\ $.  Similarly, we
define the wing-continuum boundary as the frequency shift where the
profile for wing scattering along the slab normal is equal to the
profile for continuum scattering.  From eqs. (\ref{NonResCross}) and
(\ref{ResCross}), we see that for $\Psi=0$ this boundary occurs for the
first harmonic at $|x_1| = 0.732\ E_B/E_d^1$ in the red wing and $|x_1|
= 2.73\ E_B/E_d^1$ in the blue (cf. Figure \ref{profile}c).

\medskip
The dependence of the resonant cross section on photon energy is not
strictly Lorentzian.  There is a non-Lorentzian factor, $4E^{r3}
E_B^{-1} (E^r+E_B)^{-2}$, shown in square brackets in
eq. (\ref{ResCross}).  In Figure \ref{profile}d, e, and f, we compare
$\phi_0+\phi_1$, calculated with and without the non-Lorentzian
factor, for $\mu$=0.0, 0.5, and 1.0 respectively.  The profile is
unaffected at the line center where the non-Lorentzian factor is equal
to unity.  However, this factor changes the profiles considerably in
the wings.  It raises the value of $\phi_1$ in the blue wing, but
ensures that $\phi_1 \rightarrow 0$ as $E \rightarrow 0$.  This has a
significant effect on the total profile as $\mu \rightarrow 1$, where
$\phi_0$ also goes to zero at low energies.  It is clear from the
figure that the non-Lorentzian factor can be ignored when the line
wings are optically thin.  Consequently, \cite{wang89} did not include
it in their simulations.  But, as we shall show, the non-Lorentzian
factor significantly alters the properties of radiation emerging from
line-forming regions that are optically thick in the wings.  We
therefore include it in our calculations for large optical depths.

\medskip
The code used by Wang {\it et al.} required the magnetic field to be
oriented along the slab normal ($\Psi=0$).  When this is the case, the
line-forming region is azimuthally symmetric.  Consequently, the
original code only needed to keep track of the polar angle $\theta$ of
a photon's orientation; it ignored the azimuthal angle $\varphi$.  When
$\Psi\neq 0$, the symmetry is broken, as shown in Figure
\ref{geometry}a.  Thus, in the present work, we modify the code to keep
track of both angles ($\theta_{sl}$ and $\varphi_{sl}$) and bin the
output accordingly.  However, we continue to use cross sections that
are averaged over the azimuthal angle, $\varphi$.  The
$\varphi$-dependent part of the scattering cross section is only
significant in the line wings and continuum; in the present work we
consider line-forming regions that are either not optically thick in
the wings or are azimuthally symmetric.

\medskip
\cite{lamb89} showed that relativistic kinematics has a significant
effect on the shape of the absorption profile, even in the limit
$E,kT_e \ll m_e c^2$.  We therefore use relativistic kinematics
throughout this calculation, except where we indicate otherwise.  For
zero natural line width, relativistic kinematics prohibits scattering
at the $n^{th}$ harmonic above a cutoff frequency
$E_c=(\sqrt{1+2nb}-1)\ m_e c^2/\sqrt{1-\mu^2}$ (Daugherty and Ventura
1978; Harding and Daugherty 1991; see Appendix of Wang, Wasserman, and
Lamb 1993 for a physical derivation).  \cite{ws80} show that under
physical conditions where electron recoil is important, photons escape
more readily in the red wing than in the blue.  Since there was less
of a focus on the blue wing, and for simplicity, \cite{lamb89},
\cite{wang89}, and \cite{wwl93} took the resonant scattering profile
to be zero for $E>E_c$, ignoring the effects of the finite natural
line width.  For $E<E_c$, the effects of finite line width {\it
were\/} included.  The absence of resonant scattering above $E_c(\mu)$
gave rise to a spike at small $\mu$ just blueward of the first
harmonic in some of the spectra generated by the original Monte Carlo
code. This spike contained photons which were scattered to energies
above $E_c$, either by scattering at the first harmonic or by photon
spawning due to Raman scattering at the second or third harmonics, and
which then escaped the atmosphere without further scattering (the
probability of {\it continuum\/} scattering is finite but very small
in the thin slabs used).  In the present work, we include the effect
of finite natural line width for $E > E_c$ so that the resonant
scattering profile is now small but finite above $E_c$. With this
enhancement, the spikes are smeared by scattering and no longer
appear.  However, we stress that, even though the scattering profile
is finite above the cutoff energy when the effects of natural line
width are properly treated, the profile still falls off sharply above
$E_c$, leading to a strongly asymmetric line shape at small $\mu$.

\medskip
We do not include nonresonant inverse magnetic bremsstrahlung in our
calculation. This is justified since we are interested in high photon
energies ($\gg 1$ keV) and small column depths.  Specifically, photons
with energy $E$ originating from a depth $N_{e,21} > N_{e,21}^{th}$,
where

\begin{equation}
N^{th}_{e,21} \approx 5.8\times 10^4\, \left({{kT_e}\over{5\,\,\, {\rm
                          keV} }}\right)^{1/3}\,
                          \left({{E}\over{20\,\,\, {\rm keV}
                          }}\right)^{7/6}
\label{thermdepth}
\end{equation}
(see, e.g., Nelson, Salpeter, and Wasserman 1993), will be thermalized
before escape.  For all cases we study in the present work, the column
depth $N_{e,21} < N_{e,21}^{th}$.

\bigskip
\section{Scattering Energetics and the Equilibrium Compton Temperature}
\label{secCompton}

\medskip
When a photon scatters off an electron, energy is exchanged, either
heating or cooling the atmosphere.  The Compton equilibrium temperature
$T_c$ is defined as the temperature where the heating and cooling,
summed over all scatters, balance exactly.  For media optically thin in
the continuum, the temperature is determined by resonant scattering.
The resonant Compton temperature is reached on time scales which are
short compared to most time scales of interest, such as the burst and
dynamical time scales (see Lamb, Wang, and Wasserman 1990, hereafter
LWW).  We therefore calculate spectra for atmospheres at the resonant
Compton temperature, except where we indicate otherwise.  Consequently,
it is important to understand how this temperature is affected by the
geometry of the line-forming region.

\medskip
To gain physical insight into the dependence of the resonant Compton
temperature on the field angle $\Psi$, we first calculate $T_c$
analytically in the single scattering limit (valid for line-forming
regions which are optically thin in the first harmonic).  We present
this calculation in Section \ref{Tcanalytic}.  In Section \ref{TcMC},
we present results for $T_c$ obtained from Monte Carlo simulations for
line-forming regions that are optically thick in the line core, but not
in the wings.  We discuss line-forming regions that are also optically
thick in the wings and the continuum in Section \ref{TcLarge}.

\subsection{Small Optical Depths}
\label{Tcanalytic}

\medskip
The single scattering analytic calculation of $T_c$ applies in the
limit of column depths small enough that the medium is optically thin
at the first harmonic.  Our treatment is similar but not identical to
the analytic treatment of LWW.  In their treatment, LWW used the
resonant single scattering power to calculate $T_c$. Thus, their
treatment depended solely on the distribution of injected photons with
respect to the magnetic field (i.e., on the direction of the
non-gyrating component of the electrons' velocities); it did not
address the particular geometry of the scattering medium.  In the
present analytic treatment, we explicitly consider an optically thin
slab geometry and study the dependence of $T_c$ on $\Psi$.
 In the evaluation of $T_c$, we assume an isothermal atmosphere which
is threaded by a magnetic field whose strength is much less than the
critical field $B_{\rm crit}$ and where the electron temperature $T_e$
is much less than $m_e$ (we use $\hbar=c=k=1$ throughout Section
\ref{secCompton}).

To calculate the Compton temperature, we require that the {\it net}
power in scattered photons, that is, the scattered power minus the
incident power, be equal to zero.  Equivalently, the heating and
cooling of the atmosphere by the plasma exactly balance at $T_e=T_c$
so that the total energy change of the photons $\Delta E$, which has
been summed over scatters and averaged over the scattering photon
distribution, is equal to zero.  Working in the lab frame, we have

\begin{equation}
\Delta E \equiv {{\int\,d\Omega\,dE\, n(E,\Omega)\, \,
    {\overline{\delta E}}\,N_{scat}(\tau_{To},E,\Omega)}\over
    {\int\,d\Omega\,dE\,n(E,\Omega)}}\ , \label{Dele}
\end{equation} 

\noindent
where $n(E,\Omega)\,dE\,d\Omega$ is the scattering photon number
density in the interval $(E,\Omega)$ to $(E+dE,\Omega+d\Omega)$
(assumed constant throughout the medium),
$N_{scat}(\tau_{To},E,\Omega)$ is the total number of scatters
experienced by a photon injected from position $\tau_{To}$ inside the
medium with energy $E$ and direction of propagation $\Omega$, and
${\overline{\delta E}}$ is the mean energy change per electron-photon
scattering.

Setting $\Delta E=0$ thus gives the slab Compton temperature

\begin{equation}
{T_c \over {E_B}} = {1\over{10}}\,{{ \int{{d\Omega
     }\over{\vert\mu_{sl}\vert}}\,Q(\Omega)\, (2+7\mu^2+5\mu^4)}\over{
     \int{{d\Omega }\over{\vert\mu_{sl}\vert}}\,Q(\Omega)\,
     [1+(s+2)\mu^2+(s-3)\mu^4] }}\ .
\eqnum{\ref{tceq}}
\end{equation} 

\noindent
  The derivation is given in the Appendix. In arriving at this expression
  we have assumed a separable scattering photon number density
  
\begin{equation}
n(E,\Omega) = n(E)\,Q(\Omega)
\eqnum{\ref{sep}}
\end{equation} 

\noindent  with 

\begin{equation}
s \equiv -{E\over{n(E)}}\,{{dn }\over{dE }}\Bigg\vert_{E=E_B},
\eqnum{\ref{indx}}
\end{equation}

\noindent
and used a non-relativistic one-dimensional Maxwellian for the 
electron momentum distribution along the field.

The most natural coordinate system to use to evaluate $T_c$ is the
slab coordinate system (cf. Figure \ref{geometry}), that is,
$d\Omega= d\mu_{sl}d\varphi_{sl}$, where

\begin{equation}
\mu=\mu_{sl} \cos\Psi + \sqrt{1-\mu_{sl}^2} \sin\Psi \cos\varphi_{sl}\ .
\label{musl}
\end{equation} 

\noindent
    For injection along the slab normal,
    $Q(\Omega)\propto\delta(\mu_{sl}-1)$, $\mu=\cos\Psi$, and
    eq. (\ref{tceq}) gives

\begin{equation}
{T_c \over {E_B}} = {1\over{10}}\,{{2+7\cos^2\Psi+5\cos^4\Psi }\over
                   {1+(s+2)\cos^2\Psi+(s-3)\cos^4\Psi }}\ .
\label{tcp}
\end{equation} 

\noindent
This reduces to the values found by LWW for injection along the field,

\begin{equation}
\Psi=0,\ {T_c \over E_B} = {T_c^\parallel \over E_B} \equiv \frac{7}{10s}\ ,
\label{tcpar}
\end{equation}

\noindent
and injection orthogonal to the field,

\begin{equation}
\Psi=\pi/2,\ {T_c \over E_B} ={T_c^\perp \over E_B} \equiv \frac{1}{5}\ ,
\label{tcperp}
\end{equation}

\noindent
as it must.  If the initial photon distribution covers a range of
angles, we expect the Compton temperature to fall between the two
extremes given in eqs. (\ref{tcpar}) and (\ref{tcperp}).
   
     The $1/\vert\mu_{sl}\vert$ factor in eq. (\ref{tceq}) originates
     from the slab geometry [i.e., eq. (\ref{nscat})].  It indicates
     that the dominant
     contribution comes from photons traveling at large angles to the
     slab normal, since these photons have the largest probability of
     scattering.  However eq. (\ref{tceq}) is valid only when
     $\tau/\vert\mu_{sl}\vert\ll 1$ [cf. eqs. (\ref{Delep}),
     (\ref{nscat}), and (\ref{Delef})], so the case of isotropic injection
     must be treated with some care.  For first harmonic scattering,

\begin{equation}
\tau(E=E_B,\Omega)\approx {{\tau_1}\over{\sqrt\pi}}\,{3\over4}
                 {{(1+\mu^2)}\over{\vert\mu\vert}}\ ,
\label{tauexp}
\end{equation} 

\noindent
where $\tau_1$ is given by eq. (\ref{tau1}).  For a given $\tau_1 <1$,
there is some critical $\vert\mu_{sl}^c\vert $ above which eq.
(\ref{tceq}) is valid.  Thus, for example, when $\Psi=0$ and $\tau_1\ll
1$, $\tau/\vert\mu_{sl}\vert < 1$ when $\vert\mu_{sl}\vert >
\vert\mu_{sl}^c\vert \approx (3\tau_1/4\sqrt\pi)^{1/2}$.  When
$\vert\mu_{sl}\vert<\vert\mu_{sl}^c\vert$, the single scattering
approximation fails.  When eq. (\ref{musl}) is substituted into eq.
(\ref{tceq}) and photon injection is isotropic or semi-isotropic (i.e.
isotropic over a hemisphere), the integrands in equation (\ref{tceq})
have terms of order $\mu_{sl}^{-1}$, $\mu_{sl}$, and $\mu_{sl}^3$.  If
$\tau_1$ (and therefore $\vert \mu_{sl}^c \vert$) is small enough, eq.
(\ref{tceq}) will be valid at sufficiently small values of $\mu_{sl}$
so that the $\mu_{sl}^{-1}$ terms dominate the integrals.  We thus
have, for semi-isotropic injection,

\begin{equation}
{T_c \over E_B} = {1\over {10}}\, {{16 + 28\sin^2\Psi + 15\sin^4\Psi} \over {8
     + 4(s+2)\sin^2\Psi + 3(s-3)\sin^4\Psi}}\ . \label{tci}
\end{equation} 

\noindent

       For $\Psi \to 0$, $T_c/E_B \to T_c^\perp/E_B \equiv 1/5$,
       because the dominant contribution comes from photons traveling
       nearly orthogonal to the slab normal and hence the field.  As
       $\Psi$ increases, so does $T_c$, due to the contribution of
       photons traveling at smaller angles to the field.  For $\Psi\to
       \pi/2$, $T_c/E_B \to 59/70(1+s)$ which is different from the 
       result for injection along the field ($T_c^\parallel/E_B = 7/10s$). 
         This is because, although photons with small
       $\vert\mu_{sl}\vert$ are dominant, the isotropy in $\varphi_{sl}$
       means that photons travel at all angles relative to the field.

We have calculated $T_c/E_B$ as a function of $\Psi$ from eq.
(\ref{tci}) for $s=1$.  The results are shown in Figure \ref{tcthin}.
The temperature rises from $T_c/ E_B = T_c^\perp/E_B = 1/5$ when the
field is along the slab normal ($\Psi=0$) to $T_c/E_B = 59 / 140$ when
the field is along the slab ($\Psi=\pi/2$).  Figure \ref{tcthin} also
shows Monte Carlo calculations for $B_{12}$=1.70 and $N_{e,21}=6 \times
10^{-4}$ ($\vert\mu_{sl}^c\vert\approx 0.016$).  We show the Monte
Carlo calculations for both the approximate (non-relativistic)
kinematics given by eq. (\ref{escat}) and for exact relativistic
kinematics.  The analytic model and the {\it non-relativistic} Monte
Carlo calculation show good agreement, as they should.  The value of
$T_c$ differs between the relativistic Monte Carlo calculation and the
analytic one by amounts up to $\sim$ a few times $E^r / 2 m_e$.  We
attribute the differences between the two calculations to the higher
order terms in $E^r/2 m_e$ which were not included in the kinematics in
the analytic model (eq. (\ref{escat})).  Using exact relativistic
kinematics, there are two roots for the electron momentum, $p_{min}$
and $p_{max}$ (see, e.g., Daugherty and Ventura 1978; Harding and
Daugherty 1991; Wang, Wasserman, and Lamb 1993).
This is in contrast to the single root for the non-relativistic
kinematics used in our analytic model.  Scattering via the $p_{max}$
channel involves fast moving, i.e., higher $\beta$, electrons, and is
rare for the electron temperatures we consider; typically 1-2\% of all
scatters access the $p_{max}$ channel in the relativistic Monte Carlo
simulations shown in Figure \ref{tcthin}. However, more energy is
transferred to the photons in these rare events with the net result
being that runs using relativistic kinematics including the $p_{max}$
channel typically had more efficient electron cooling and hence lower
$T_c/E_B$ (for given $\Psi$) than runs without this channel. This is
evident in Figure \ref{tcthin}.

\subsection{Moderate Optical Depths}
\label{TcMC}

\medskip
At larger column depths, $\tau/\vert \mu_{sl} \vert > 1$ and
eq. (\ref{nscat}) is no longer valid.  Consequently, we need to
consider the effects of multiple scatterings and we can not use our
analytic treatment to calculate $T_c$.  Instead we use our Monte Carlo
code to calculate the emerging photon spectrum and determine the amount
of energy transferred to the photons while passing through the slab.
By varying the electron temperature we can determine the temperature at
which the net energy transferred is zero. Using this technique, LWW
calculate the Compton temperature for semi-isotropic injection in a
line-forming region with a 1-0 geometry and $\Psi=0$, column depths in
the range $0.12 \le N_{e,21} \le 12$, and magnetic fields in the range
$1.50 \le B_{12} \le 2.10$.  They find that $T_c/E_B \approx 0.27$.  In
the ranges considered, they find the Compton temperature to be
relatively insensitive to the magnetic field and to decrease slightly
with increasing column depth.

\medskip
In the present work, we fix the column depth and magnetic field and
study the effects of the source plane position and the field
orientation, as well as the effects of the higher harmonics and the
natural line width.  We use $N_{e,21}$=1.2, $B_{12}$=1.70, and a power
law injected spectrum with $s=$1.  Figure \ref{tcthick} shows the
Compton temperature $T_c/E_B$ as a function of the field orientation
for both the 1-0 and 1-1 geometries.  The temperatures are generally
higher than in the optically thin case.  The increase is most dramatic
when the field is parallel to the slab normal ($\Psi=0$).  To
understand this, recall that in the optically thin case, almost all
photons that scatter are moving perpendicular to the slab normal, and
therefore have $\mu=\mu_{sl}\approx 0$ when $\Psi=0$.  However, in the
optically thick case, there is an increase in the number of photons
moving at larger values of $\mu$ before scattering, which raises the
Compton temperature (cf. eqs. [\ref{tcpar}], [\ref{tcperp}]).  The
increase in high $\mu$ photons that scatter is due to the increased
optical depth for photons moving parallel to the field (along the slab
normal), and the multiple scattering of photons that were injected
perpendicular to the field (along the slab).

\medskip
While the analytic model assumes zero natural line width and includes
first harmonic scattering only, the calculations shown in Figure
\ref{tcthick} assume finite natural line width and include scattering
at the first three harmonics.  As shown in table \ref{Ttemps}, the
effect of the finite line width and higher harmonics is small.

\medskip
As we increase the column depth to $N_{e,21}$ = 12, we find a small
{\it increase} in Compton temperature, in contrast to the results of
LWW.  The difference is due to the effects of blue wing scattering.
Because higher energy photons are able to scatter, the cooling of the
electrons is less efficient, which raises the Compton temperature.

\subsection{Large Optical Depths}
\label{TcLarge}
In a closed system in thermal equilibrium, in which the photon density
is low enough that stimulated scattering can be ignored, the photons
have a Wien distribution with temperature $T_\gamma=T_e$ (see, e.g.
Rybicki and Lightman 1979).  There is no net energy transfer between
the photons and the electrons, and the photons have an isotropic
angular distribution, i.e. $N(E)=0$, where $N(E)$ is the net photon
flux.  We expect that at optical depths that are large enough so that
$\tau \gg 1$ and the flux to density ratio, $N(E)/(c n(E)) \ll 1$, the
photons behave approximately as in a closed system.

\medskip
In section \ref{TcMC} we discussed scattering energetics in
line-forming regions that are optically thick in the line core and
optically thin in the wings and the continuum.  In this section we
examine the effects of larger optical depths by considering
line-forming regions with $\Psi=0$, $B_{12}=3.5$, and
$N_{e,21}=15,000\ (\tau_{To}=10)$.  The line-forming region is
optically thick in both the line wings ($a\tau_1 \gg 1$) and the
continuum ($\tau_{To} \gg 1$).  For comparison, we also consider a line
forming region with $N_{e,21}=1,500\ (\tau_{To}=1)$, which is optically
thick in the wings but only marginally thick in the continuum, and a
line-forming region with $N_{e,21}=30\ (\tau_{To}=0.02)$ which is
marginally optically thick in the wings and optically thin in the
continuum.  All three line-forming regions are optically thick in the
line core ($\tau_1 \gg 1$).  The key parameters for the runs are listed
in Table \ref{Tparms}.  In addition, the mean number of scatters
experienced per escaping photon, $N_{scat}$, and the mean energy of
escaping photons (averaged over the emerging photon distribution; see
below) are listed.

\medskip
To simulate the behavior of a system with large optical depth, we
inject a Wien spectrum with $T_\gamma=T_e$.  For $T_e$, we adopt a
fiducial value of $kT_e/E_B=1/4$, which is approximately the Compton
temperature for resonant scattering (cf. Section \ref{TcMC}).  We
include only continuum and first harmonic scattering in these
calculations.  Higher harmonics are not important, due to the
exponential rollover in the continuum spectrum which occurs well below
the first harmonic. Thus, eq. (\ref{BasicAssump}) with only the first
term on the right hand side is a valid approximation for these
simulations, even for line-forming regions that are optically thick in
the line wings.

\medskip
As Table \ref{Tparms} shows, at the largest optical depths, the mean
energy of the transmitted photons is considerably smaller than the
mean injected energy.  For example, for $N_{e,21}=15,000\
(\tau_{To}=10)$ and the 1-1 geometry, the mean energy of the
transmitted photons is 15.5 keV, compared to a mean energy of
$3kT_\gamma=30$ keV for the injected photons.  Most of the energy
redistribution occurs deep in the slab, as shown in Figure
\ref{netflux}.  Here we plot $\langle E_t(\tau_T)\rangle$, the mean
photon energy associated with the net (upward) flux crossing a given
plane inside the slab and is calculated from

\begin{equation}
 \langle E_t(\tau_T)\rangle = {{\int_0^\infty dE\, E\,[N_{\rm up}(E,\tau_T)-
                           N_{\rm down}(E,\tau_T)]}\over {\int_0^\infty dE\,
                           [ N_{\rm up}(E,\tau_T)-N_{\rm down}(E,\tau_T) ].}}
\label{netfluxdef}
\end{equation}

\noindent
  At the edge of the atmosphere, $N_{\rm down}(E,\tau_T=0)\equiv0$, 
  and $\langle E_t(\tau_T=0)\rangle$ becomes simply the mean energy of escaping 
  photons averaged over the emergent photon distribution (cf. last column in 
  Table \ref{Tparms}).

\medskip
It is evident from this figure that an optical depth of $\tau_T=10$ is
not sufficient for thermal equilibrium between the electrons and the
photons.  On the contrary, as the photons move upwards they steadily
lose energy to the electrons until they reach a depth of about
$\tau_T=7$.  Above this depth, the shape of the spectrum is altered by
multiple scattering, but the mean energy of the distribution remains
the same until the photons get close to the surface.  Above $\tau_T
\approx 0.004$, the line wings are optically thin and there is a slight
increase in $\langle E_t(\tau_T)\rangle$.

\medskip
The reason that thermal equilibrium does not occur at $\tau_T=10$ is
apparent in eq. (\ref{NonResCross}) and Figure \ref{profile}c.  For $E
\ll E_B$, the scattering profile for a $\mu=1$ photon decreases with
energy like $(E/E_B)^2$.  Thus, at {\it any} depth $\tau_T$, there
exists some energy, $E_{\rm thin}(\tau_T)$, such that the plasma
appears optically thin to $\mu=1$ photons with $E<E_{\rm
thin}(\tau_T)$.  If $\tau_T$ is large enough so that $E_{\rm thin}\ll
E_B$, then from eqs.  (\ref{ClassCross})-(\ref{ResCross}),

\begin{equation}
{ {E_{\rm thin}(\tau_T)}\over {E_B} } =\left ( {2+\tau_T+\sqrt{8\tau_T+\tau_T^2} \over 2}
\right )^{-1/2}.
\end{equation}

\noindent
Because of the strong angle and frequency dependence of the cross
section in a strong magnetic field, the use of the Thomson depth to
indicate optical depth in the continuum is misleading.  Even at
$\tau_T=10$, a significant number of photons can escape without
scattering.

\medskip
The energy shift in Figure \ref{netflux} occurs because photons
diffuse in angle- and frequency-space until their angle with the
magnetic field is small and $E \lesssim E_{\rm thin}(\tau_T)$.  They
are then able to escape without further scattering.  The importance of
this redistribution is illustrated by comparing the cumulative
spectrum of the emerging photons with that of the injected photons, as
shown in Figure \ref{cumspec}.  As the figure shows, even though only
14\% of the photons are injected with $E<E_{\rm thin}(7)=12.9$ keV,
45\% escape below this energy.  This effect could be even more
pronounced if we did not use polarization-averaged cross sections; at
low energies, the scattering profile for extraordinary mode photons
goes like $(E/E_B)^2$, regardless of the direction they are
traveling.

\medskip
To test the hypothesis that the shift in the photon energies is due to
the form of the scattering cross section at low $E$, we ran a 1-1,
$N_{e,21}=15,000\ (\tau_{To}=10)$ simulation in which the
non-Lorentzian factor was removed from the scattering cross section.
As Figure \ref{profile}f shows, without this factor, the scattering
profile approaches $\sigma_T$ as $E \rightarrow 0$, so the
line-forming region is no longer optically thin to low energy photons.
As shown in Table \ref{Tparms}, $\langle E_t(0)\rangle=24.7$ keV in
this simulation, much larger than in the simulation with the
non-Lorentzian factor included.  In addition, without the
non-Lorentzian factor, the ratio $N(E)/(c n(E))$ remains much smaller
than unity through most of the slab, as expected for a thermal
distribution of photons.  Clearly the detailed behavior of the
scattering cross section plays an important role in forming the
redshifted spectrum of the escaping photons.

\medskip
The transfer of energy from the photons to the electrons deep in the
slab implies that the electron temperature $T_e=E_B/4$ assumed in the
calculation is less than the Compton temperature at large depths.  In
this situation, the photons will heat the electrons until $T_C$ is
reached.  However, as Figure (\ref{netflux}) indicates, $T_e$ already
equals $T_C$ above $\tau_T \approx 7$.  The shape of the curve
indicates that a self-consistent calculation requires a temperature
profile, $T_e(\tau_T)$, rather than a constant value of $T_e$
throughout the slab.  This result is consistent with the finding of
\cite{bulik93} that $T_e$ rises slightly below $\tau_T \sim$ a few (see
also Nagel 1981; Miller, Wasserman, and Salpeter 1989).  The temperature
profile could be implemented by dividing the slab into several zones
with different electron temperatures, as suggested by LWW.

\bigskip
\section{Monte Carlo Spectra}
\label{secSpectra}

\medskip
Our Monte Carlo simulations reveal a rich variety of line properties
which vary with the viewing angle, the magnetic field orientation, the
optical depth, and the slab geometry.  For line-forming regions that
are optically thick in the line core, but optically thin in the line
wings and the continuum, we discuss the line shapes in Section
\ref{ssecShapes} and the equivalent widths in Section \ref{ssecEW}.
These simulations reveal prominent shoulders on both sides of the
first harmonic line when the radiation is viewed at $\mu_{sl} \gtrsim
0.25$, regardless of the field orientation.  We discuss the line
shoulders in Section \ref{ssecShoulders}.  In Section \ref{ssecBigTau}
we consider line-forming regions that are optically thick in the
line wings and the continuum, in addition to the core.

\subsection{Line Shapes for Moderate Optical Depths}
\label{ssecShapes}

\medskip
Figures \ref{spectra0}, \ref{spectra45} and \ref{spectra90} show Monte
Carlo scattering spectra for $\Psi =0$, $\Psi=\pi/4$, and $\Psi
=\pi/2$, respectively.  In all three figures, the column depth is
$N_{e,21}=1.2$, which is optically thick in the line core and
optically thin in the wings.  We include results for both the 1-0 and
1-1 geometries and for several viewing angles.  To produce the
$\Psi=0$ spectra, we inject a total of one million photons into the
slab isotropically and record the photons emerging from the slab in
one of eight bins in $\mu_{sl}$ ($0<\mu_{sl}\leq 1$).  To produce the
$\Psi \neq 0$ spectra, we inject eight million photons and record the
emerging photons in 80 angular bins (10 bins in $\varphi_{sl}$ between
0 and $\pi$ for each of 8 bins in $\mu_{sl}$).  In each panel the 1-1
spectrum is normalized to have unit area and the 1-0 and pure
absorption spectra are normalized to match the 1-1 spectrum in the
continuum.

\medskip
For comparison, we also show pure absorption spectra,

\begin{equation}
N_{abs}(E,\Omega_{sl}) = N_i(E) \exp\left ( - {N_e \sigma_T \phi \over
|\mu_{sl}|}\right )\ ,
\label{nabs}
\end{equation}

\noindent
where the scattering profile $\phi$ is given by eq.
(\ref{ScatProfile}).  $N_{abs}(E,\Omega_{sl})$ is the spectrum in
which photons that have undergone the absorption that initiates a
resonant scatter are not re-emitted.  We can explain many of the
properties of the spectra in terms of the scattering profile and the
geometry of the line-forming region.

\medskip
The second and third harmonics have shapes similar to absorption lines
because most of the photons that scatter at these energies are Raman
scattered --- i.e., they are absorbed and then re-emitted as two or
three lower harmonic photons.  The line properties of the higher
harmonic features can therefore be understood in terms of eq.
(\ref{nabs}) and the scattering profiles.  To illustrate this, we
combine eqs. (\ref{sigrest}), (\ref{siglab}), (\ref{NScatProfile}), and
(\ref{max}) and let $\Gamma_n \rightarrow 0$ and $\beta \ll 1$ to get
the non-relativistic resonant scattering profile for the $n^{th}$
harmonic with zero natural line width,

\begin{equation}
\phi_n \approx {3 \over 8}
{\pi \over \alpha} b^{n-1} {(n^2/2)^{n-1} \over (n-1)!} (1 + \mu^{r2}
) (1 - \mu^{r2})^{n-1} { \exp \left [ -({E-E_n \over E_d^n \mu})^2\right
] \over \sqrt{\pi} 
   E_d^n |\mu|}\ .
\label{philab}
\end{equation} 

\noindent
It is evident from eqs. (\ref{philab}) and (\ref{nabs}) that while the
Doppler factor $E_d^n |\mu|$ broadens the lines when viewed {\it along
the field}, the line-of-sight column depth $N_e/|\mu_{sl}|$ deepens the
lines when viewed {\it along the slab}.  Thus, when the magnetic
field is parallel to the slab normal ($\Psi=0$), the lines become
deeper and narrower as the viewing angle moves from $\mu_{sl}=1$ to
$\mu_{sl}=0$ (Figure \ref{spectra0}).  When the field is perpendicular
to the slab normal however ($\Psi=\pi/2$), these two effects combine to
provide especially broad and deep lines when viewed perpendicular to
the slab normal and at modest angles to the magnetic field (e.g.,
Figure \ref{spectra90}k).  When the viewing angle is directly {\it
along} the field (e.g., Figures \ref{spectra0}a and \ref{spectra90}j),
the scattering profile for the higher harmonics is dominated by the
$(1-\mu^{r2})^{n-1}$ factor and the higher harmonics are suppressed.

\medskip
The properties of the first harmonic are determined by multiple
scatterings and photon spawning.  Consequently, there is no simple
analytic expression that describes the first harmonic scattering line.
However, though the physics of line-formation are fundamentally
different, the full-width half-maximum of the first harmonic {\it
scattering} line tends to be similar to that of the first harmonic {\it
absorption} line.  Thus, eqs. (\ref{nabs}) and (\ref{philab}) do
provide some insight into the first harmonic.  Up to a logarithmic
factor $\sim (\ln\tau_1)^{1/2}\sim 1$, the Doppler width dominates the
width of the first harmonic line. This is because $\delta E/E\ll 1$ in
core scattering and there is negligible frequency redistribution in the
line wings for the cases we study in Figures
(\ref{spectra0})--(\ref{spectra90}).

\subsection{Line Equivalent Widths for Moderate Optical Depths}
\label{ssecEW}

\medskip
The equivalent width of the first harmonic line is plotted as a
function of viewing angle in Figures \ref{ewa}, \ref{ewb}, and
\ref{ewc}.  In calculating the first harmonic equivalent width, we need
to adjust for any overlap between the first and second harmonic lines
in the emergent spectra.  For column depths $N_{e,21}/|\mu_{sl}|
\gtrsim 6 (kT_e/ 1\ {\rm keV})$, the line wings of the first harmonic
can have optical depths $\gtrsim 1$ so frequency redistribution in the
line wings can be significant.  Thus, at large values of $N_{e,21}$ or
small values of $|\mu_{sl}|$, the first and second harmonics begin to
overlap (see, e.g., Figure \ref{spectra90}j). To correct for this, we
assume that the second harmonic is approximately an absorption line and
we calculate the equivalent width of the first harmonic according to:
\medskip

\begin{equation}
W_{E1}(\Omega_{sl}) = \int {   {N_i(E) g_2(E,\Omega_{sl}) -
          N(E,\Omega_{sl}) } \over { N_i(E) g_2(E,\Omega_{sl})  } } dE
\end{equation}

\medskip
\noindent 
where $N(E,\Omega_{sl})$ is the transmitted spectrum and
$g_2(E,\Omega_{sl})$ is as defined in eq. (\ref{nabs}) with
$\phi\to\phi_2$.  We stress that this formula {\it assumes\/} the
spectrum $N(E,\Omega_{sl})$ is accurate and merely corrects for the
overlap in the calculation of $W_{E1}$.  It {\it does not\/} correct
for the inaccuracy in our modeling of the transfer through a scattering
medium that could be optically thick in the line wings (cf. eq.
\ref{BasicAssump} and associated discussion).

\medskip
We display $W_{E1}$ for slabs with $\Psi=\pi/2$ in Figure \ref{ewa} for
the 1-0 geometry and in Figure \ref{ewb} for the 1-1 geometry.
$B_{12}=1.7$ and $N_{e,21}=1.2$ in both figures.  For viewing along the
slab (small $\mu_{sl}$), the equivalent
 width decreases as the azimuth $\varphi_{sl}$ moves from 0 to
 $\pi/2$.  This is because the scattering profile for the first
harmonic, unlike the higher harmonics, has no $1-\mu^{r2}$ factor that
causes the cross section to vanish along the field.  Photons
propagating along the field see a larger cross section and Doppler
width which tends to scatter them out of the line of sight, while
photons propagating orthogonal to the field see a reduced scattering
cross section and Doppler width which reduces the chance of
scattering.  As $\mu_{sl}$ increases, the equivalent width decreases
because photons escape more easily in directions transverse to the slab
where the column depth is lower.

\medskip
We display $W_{E1}$ for slabs with $\Psi=0$ and compare the results
with those for $\Psi=\pi/2$ in Figure \ref{ewc}.  The figure also
illustrates the dependence of equivalent width ($W_{E1}$) on column
depth.  The figure shows $W_{E1}$ as a function of $\mu_{sl}$ for
$B_{12}$=1.7 for $N_{e,21}$=0.12, 1.2, and 12; and for both the 1-0 and
1-1 geometries.  The azimuthal angle is $0<\varphi_{sl}<\pi/8$ for
$\Psi=\pi/2$ and $0<\varphi_{sl}<2\pi$ for $\Psi=0$.  The figure shows
that the geometry can have a larger effect on the equivalent width than
the column depth does.  For all geometries, the equivalent widths are
largest when the spectrum is viewed along the slab ($\mu_{sl}
\rightarrow 0$).  In the 1-0 geometry with $\Psi=0$, increasing the
column depth two orders of magnitude from $N_{e,21}$=0.12 to 12
increases the width at $0<\mu_{sl}<1/8$ from $W_{E1}/E_B=0.096$ to
0.39. However, keeping $N_{e,21}$ at 0.12 but rotating the field so it
lies along the slab increases the equivalent width in this $\mu_{sl}$
bin from $W_{E1}/E_B=0.096$ to 0.64.

\medskip
As $\mu_{sl}$ approaches unity, the equivalent width decreases and can
become negative. This is especially true for the 1-1 geometry. A
negative equivalent width corresponds to an emission-like feature.  The
presence of such a feature in an angular bin requires a surplus of
photons emerging in the bin compared with the number of photons that
were injected.  In Figures \ref{spectra0}-\ref{spectra90}, such
features appear as shoulders on either side of the line center.  The
shoulders are very prominent in the 1-1 geometry, but less so in the
1-0.  For example, as we see in Figure \ref{ewc}d, $W_{E1}/E_B$=--0.96
in the $7/8<\mu_{sl}<1$ bin in the 1-1 geometry, but only --0.066 in
the 1-0.  Physically, these shoulders are the result of angular and
frequency redistribution in electron-photon resonant scattering.

\subsection{Line Shoulders}
\label{ssecShoulders}

\medskip
Many authors discuss line shoulders.  \cite{ws80} predict that for
physical conditions where electron recoil is important (e.g.,
accretion-powered pulsar line-forming regions), there is an excess of
photons escaping in the red wing compared with the blue.  \cite{am89},
\cite{ne92}, and \cite{nishimura94} report shoulders in spectra
generated by Feautrier calculations with 1-0 geometries and $N_{e,21}$
varying from $\sim 0.1$ to $\sim 100$.   The shoulders reported by
\cite{am89} are small, in agreement with our 1-0 results.  In contrast,
\cite{nishimura94} finds very large shoulders with $W_{E1}/E_B<-10$ in
some cases.  Further, he reports that the shoulders are most prominent
at {\it small} $\mu_{sl}$, in direct contradiction to our results.
\cite{ah96} also report shoulders, which they observe in spectra
generated by a Monte Carlo code for both 1-1 and cylindrical geometries
with column depths up to $N_{e,21} \sim 5$.  In this section, we
develop a better understanding of line shoulders by investigating how
they are affected by viewing angle, spawning, electron temperature, and
source plane position.

\medskip
\cite{chandrasekhar60} shows that photons that scatter isotropically in
a slab atmosphere emerge disproportionately at high $\mu_{sl}$
because of the lower column depth along the slab normal.  The tendency
of photons to emerge at high $\mu_{sl}$ is the reason the shoulders in
the present calculation become more prominent as $\mu_{sl} \rightarrow
1$.  The shoulders are weak or nonexistent for low $\mu_{sl}$ because
the enhanced column depth along the slab ensures that most photons are
scattered out of the line of sight, resulting in the formation of
absorption-like features.  Figure \ref{binhist} displays Monte Carlo
calculations of the emergent angular distribution of resonantly
scattered photons. The photons are injected monochromatically at
$E=E_B$ so that every photon scatters.
   As the figure shows, the number of photons {\it emerging\/} generally
   increases with $\mu_{sl}$, even though they were {\it injected\/}
   isotropically. This is true for both the 1-0 and 1-1 geometries and
   for fields both parallel and perpendicular to the slab normal. 
In the 1-0 geometry (Figures \ref{binhist}a and c), there is an excess
of reflected photons over transmitted photons because of the shorter
path length for the former.  This contrasts with the 1-1 geometry
(Figures \ref{binhist}b and d), where the line forming region is
symmetric about the source plane.  There is no short escape path and
the transmitted and reflected spectra are (by symmetry) identical.
    In both geometries, few photons escape
    along the slab owing to the large path length.
    In the $\Psi=0$ case, escape along the slab normal is favored 
    by both the short path length and by a scattering cross
    section which is largest for a scattered photon direction
$\mu_s = 1$.   The emergent      
    angular distribution peaks at $\mu_{sl}=1$ (cf. Figures 
    \ref{binhist}a,b). By contrast, in the $\Psi=\pi/2$ case, 
    escape along the slab normal, while favored by the shorter
    path length, is discouraged by the scattering cross section. 
The emergent angular distribution is, therefore, peaked 
     in a direction determined by a compromise between the most favored 
     scattered angle ($\mu_{sl}\to 0$) and the shortest path length 
     ($\mu_{sl}\to 1$) (cf. Figures \ref{binhist}c and d).
 
Figures \ref{shoulders0} and \ref{shoulders90} show spectra emerging
from a line-forming region with a column depth $N_{e,21}=1.2$ threaded
by a field of strength $B_{12} = 1.7$ with $\Psi=0$ and $\pi/2$,
respectively.  The injected photon spectrum is $\propto 1/E$.  When the
viewing angle is perpendicular to the slab normal, the scattered
spectrum is similar to a pure absorption spectrum. When the viewing
angle is parallel to the slab normal, shoulders can appear.  It is
evident from Figures \ref{shoulders0}a and b and \ref{shoulders90}a and
b that spawning enhances the shoulders by providing an additional
source of first harmonic photons. However, the shoulders are prominent
in the 1-1 geometry even when spawning is not included in the
calculation.  This is because photons injected initially into the
bottom half of the slab in the 1-1 geometry can escape through the top
half after multiple scatters thereby providing an effective additional
source of first harmonic photons (see Figure \ref{flux0} below and
associated discussion).

\medskip
Figures \ref{shoulders0} and \ref{shoulders90} also reveal that the
spacing of the line shoulders is of the same order as the width of the
corresponding pure absorption lines.  This simply reflects the fact
that in media optically thin in the line wings the shoulder spacing is
given approximately by the Doppler width $2E_d^1\vert\mu^{esc}\vert$
times a factor $\sim(\ln\tau_1)^{1/2}\sim 1$ as a result of multiple
scatters (cf. Osterbrock 1962; Wang, Wasserman, Salpeter 1988).
 
\medskip
The relative size of the red and blue shoulders is determined by the
relationship between the electron temperature and the Compton
temperature.  At the Compton temperature, by definition, the photons
on average lose as much energy as they gain.  We therefore expect the
area under the two shoulders in the photon {\it energy} spectrum to be
approximately equal when $T_e=T_c$.  

\noindent
As we see in Figures \ref{shoulders_vs_t}c and d, this is the case for
a 1-1 slab atmosphere with $B_{12}=1.7$, $N_{e,21}=1.2$, $\Psi=0$ and
$\pi/2$, and a 1/E injected photon number spectrum.  In order to
illustrate the total change in energy of the photons, we use
angle-integrated spectra in this figure.  For $T_e<T_c$, the photons on
average lose more energy than they gain and the red shoulder is larger
(Figures \ref{shoulders_vs_t}a and b).  This result is consistent with
\cite{am89}, who use a temperature (5.2 keV) that is below the Compton
temperature, and report a slightly greater flux in the red shoulder
over the blue.  For $T_e > T_c$, we find the blue shoulder is larger
than the red, as expected (Figures \ref{shoulders_vs_t}e and f).

\medskip
The line shoulders are more prominent in the 1-1 geometry than in the
1-0 at modest optical depths because of an excess in the flux of
photons near the line center in the 1-1.  We illustrate this for the
case of $\Psi=0$, $B_{12}=1.7$, and $N_{e,21}=1.2$. Figure \ref{flux0}
shows the angle-integrated flux of photons moving upwards
($\mu_{sl}>0$) as a function of energy at four points in the slab:
$\tau_T/\tau_{To}$=1.00 (the source plane), 0.32, 0.02, and 0.00 (the
upper surface).  At the source plane, the excess of photons in the 1-1
geometry takes the form of a prominent peak (Figure \ref{flux0}d).
The reason for the peak is that as line photons scatter, they can
cross the source plane many times.  The peak contains photons which
were injected towards the bottom of the slab ($\mu_{inj}<0$), but
which subsequently scattered upwards and crossed the source plane (an
odd number of times, in general).  The peak also contains photons
which were injected towards the top of the slab ($\mu_{inj}>0$), but
which, in the course of multiple scatters cross the source plane twice
(or, in general, an even number of times) --- once downwards and once
upwards.  In other words, the photon flux in the 1-1 geometry, like
the emergent spectrum, is the sum of transmitted and reflected
components (cf. Section \ref{seclfr}).  The two components contribute
about equally to the peak.  The peak does not appear in the 1-0
geometry because photons that are traveling downwards from the source
plane have escaped the slab and are part of the reflected spectrum.

The peak is responsible for the prominence of the shoulders in the 1-1
geometry.  As the photons in the peak move upwards through the slab
their frequencies are redistributed in multiple scatterings --- they
are scattered out of the line --- and shoulders form.  Shoulders are
diminished in the 1-0 geometry, compared to their appearance in the
1-1, because there is no photon excess near the line center.  Note also
the formation of the higher harmonic features, which, being effectively
true absorption features, is essentially independent of geometry.

\subsection{Large Optical Depths}
\label{ssecBigTau}

\medskip 
In sections \ref{ssecShapes}--\ref{ssecShoulders} we discussed the
spectra emerging from line forming regions with a moderate column
depth of $N_{e,21}=1.2\ (\tau_{To}=0.0008)$.  But the emission regions
of accretion-powered pulsars are thought to have column depths
$N_{e,21} \sim 10^3 - 10^4$.  In this section we discuss our Monte
Carlo results for the emerging spectra for slabs with column depths in
this range.  
Table \ref{Tparms}, above, lists the key parameters for these simulations.  

\medskip
Figure \ref{flux} shows the angle-integrated flux of photons moving upwards 
($\mu_{sl}>0$) as a function of energy at four points in the slab: 
$\tau_T/\tau_{To}$=1.00 (the source plane), 0.32, 0.02, and 0.00 (the upper surface).
As expected, the line becomes broader and deeper as the column depth increases. 
At $N_{e,21}=15,000$, virtually no photons escape in the line core.

\medskip
The $N_{e,21}=30$ simulation shows a prominent red wing shoulder. The
blue shoulder is less evident, due in part to the exponential decline
of the injected spectrum above the line energy.  At $N_{e,21}=1,500$,
substantial scattering in the wings significantly shifts the centroids
of the shoulders and broadens them.  At $N_{e,21}=15,000$, the slab is
optically thick in the continuum.  Scattering in the continuum broadens
the shoulders so that they are no longer discernible.

\medskip
The influence of the continuum can be seen in the mean number of
scatters a photon experiences before escaping the slab.  Taking a
typical $\mu\sim0.5$, we see from Figure \ref{profile} that continuum
scattering dominates below $\sim 20$ keV.  For $N_{e,21}=30$ and
1,500, the mean energy of escaping photons is about 29 and 25 keV,
respectively, so that photons escape primarily in the wings after
multiple core-wing excursions (though for $N_{e,21}=30$, the wings are
only marginally thick). The number of scatters, $N_{scat}$, is then
dominated by core and wing scatters.  Since these scatters are
accumulated mostly in the region where $T_e\sim T_C$ (cf. Figure
\ref{netflux}), $N_{scat}$ follows the zero recoil scaling, viz,
$N_{scat}\propto\tau_1$ (Adams 1971; Wasserman \& Salpeter 1980).  For
$N_{e,21}=15,000$, the mean energy of escaping photons is about 15
keV, so that photons escape primarily in the continuum.  In this case,
photons enter the continuum after multiple core-wing excursions.
Since the continuum is optically thick, they can return to the wings
after many scatters in the continuum and escape in the continuum after
a few such continuum-wing transitions. The addition of the continuum
domain breaks the $N_{scat}\propto \tau_1$ scaling and increases
$N_{scat}$ above that expected for pure line transfer.  This increase
can be seen clearly in Table \ref{Tparms} where $N_{scat}$ for runs
both with and without the continuum are listed.  For $\tau_{To} = 10$,
core and wing scatters still dominate the total number of scatters so
that $N_{scat}\propto \tau_1$ remains good to $\lesssim 20$\%.  For
much larger column depths, however, photons escape in the continuum
only after many continuum-wing-core transitions, thereby resulting in
much larger $N_{scat}$. Thus, at $\tau_{To}=100$, for instance, we
find $N_{scat}\approx6\times10^6$ with the continuum turned on and
about $1\times10^6$ with it turned off. The mean energy of escaping
photons in this case is about 10 keV.  The number of (line) scatters
with continuum turned off ($\sim 10^6$) is less than that expected
from the $N_{scat}\propto \tau_1$ scaling ($\sim 2\times10^6$) because
this scaling ignores the non-Lorentzian factor
(cf. eq. [\ref{ResCross}]) which substantially reduces the scattering
cross section at very low energies.

\subsection{Cylindrical Geometry}

\medskip
If the radiation from an accretion-powered pulsar emerges from the
stellar surface, the spectrum is similar to the spectrum emerging from
a 1-1 line-forming region with $\Psi=0$.  If the spectrum emerges
from the sides of a cylindrical accretion column, however,  it will be
similar to that of a 1-1 line-forming region with $\Psi=\pi/2$ viewed
in the $\varphi_{sl}\sim 0$ plane.  We illustrate the latter
case in Figure \ref{cyl}, which shows a Monte Carlo spectrum
emerging from a cylindrical line-forming region.  The magnetic field is
oriented along the cylinder axis and has strength $B_{12}=1.7$.  The
column depth, measured radially from the cylinder axis to the surface
is $N_{e,21}$=1.2.  A power law spectrum with $s=1$ is injected along
the cylinder axis.  The figure shows that the cylindrical spectrum is
qualitatively similar to a 1-1 spectrum with $\Psi=\pi/2$ viewed in the
$\varphi_{sl}\sim 0$ plane.  Shoulders appear in the spectrum at
small values of $\mu$ (Figure \ref{cyl}a) but not large values (Figure
\ref{cyl}d).  The separation between the shoulder peaks is comparable
to the line width for an absorption spectrum and increases as the line
of sight approaches the cylinder axis due to the increase in the
Doppler width.  We stress that the typical radial column depth expected
for an accretion column is $N_{e,21} \sim 10^3-10^4$, much larger than
the value used in Figure \ref{cyl}.  But we expect that, as the column
depth of a cylindrical line-forming region increases, the properties of
the emerging spectrum continue to be similar to those of a 1-1
line-forming region with $\Psi=\pi/2$, $\varphi_{sl}\sim 0$, and
similar column depth.  Specifically, the spectrum emerging from a
cylindrical line-forming region with high $N_e$ should possess a
cyclotron line with large equivalent width at all viewing angles, and
with no visible shoulders.

\section{Implications for Accretion-Powered Pulsars and Gamma-Ray Bursters}
\label{secDisc}

Our understanding of how cyclotron line properties depend on geometry
can provide important insights into the sources of observed cyclotron
lines.  It can shed light on the optical depth and location of the
line-forming region in accretion powered pulsars; for example, whether
line-formation occurs in the accretion column or in a thin scattering
layer located in the pulsar magnetosphere.  Understanding geometrical
effects can also explain how cyclotron lines in gamma-ray bursts could
be formed, e.g., by sources in a galactic corona.

\subsection{Accretion-Powered Pulsars}
\label{ssecAPP}

\medskip
Constructing fully self-consistent accretion-powered pulsar models that
produce spectra in agreement with observation presents many
challenges.  These include determining what spectrum is appropriate for
the injected photons, explaining why observed accretion-powered pulsar
spectra contain an excess of photons at low energies compared with a
Wien function, and reproducing the properties of the observed cyclotron
features.

\medskip
If the luminosity of the pulsar is small enough so that radiation
forces are not important, the accreting material is stopped and the
photons are produced at depths $N_{e,21} \sim 15,000-150,000\ (\tau_T
\sim 10-100)$ beneath the stellar surface (Miller, Salpeter, and
Wasserman 1987; Miller, Wasserman, and Salpeter 1989).  Since the
stopping depth is optically thick to (non-magnetic) Thomson scattering,
many calculations assume a Wien spectrum at these depths.  However, as
we discussed in section \ref{TcLarge}, in a strong magnetic field the
photons will not generally have a Wien distribution at these depths.
Consequently, the injected spectrum needs to reflect the physics of the
stopping process.

\medskip
The $(E/E_B)^2$ dependence of the cross section for $E \ll E_B$ may
play an important role in creating the photon excess at low energies.
The excess in our own simulations is small compared with observations,
but this is most likely due to our choice of a Wien distribution for
the injected photons and our neglect of the velocity of the accreting
material above the stopping depth.

\medskip 
Observed cyclotron features are broad, but shallow.  For example,
\cite{soong90} report a full-width half-maximum of 15.8 keV, but an
equivalent width of only 10.3 keV for the line in the phase-averaged
spectrum of Her X-1 (centroid at $\sim 35$ keV).  In
contrast, as we saw in Figure \ref{flux}, theoretical calculations for
line-forming regions with column depths typical of accretion columns
($N_{e,21} \sim 10^3-10^4$; Lamb, Pethick, and Pines 1973) generate
lines which are extremely deep, with large equivalent widths.
Similarly, in fits to the spectra of 4U 1538-52 and Vela X-1, Bulik
{\it et al.}  (1992, 1995) are unable to generate sufficiently broad
lines in models with constant field strength.  However, using a model
with two field components, differing in strength by a factor $\sim 5$,
they obtain acceptable fits to the data.  They suggest that scattering
at the lower field component could correspond to scattering in a dipole
field, $B(r) \sim B_o (R/r)^3$ at an altitude $r \gtrsim 0.3 R$, where
$B_o$ is the surface dipole field strength and $R$ is the stellar
radius. The scattering could occur in an accretion mound, as in the
model of \cite{bak91}, or in a suspended scattering atmosphere, as in
the model of \cite{ds91} and \cite{sd94}.  Thus the spectrum could be
the product of a combination of scattering near the surface and
scattering in the magnetosphere.

\medskip
The spectral signature of the geometry of the line-forming regions can
provide insight into the contribution of each region to the spectrum.
At moderate optical depths, spectra formed in the 1-1 geometry are
characterized by prominent shoulders; these are smaller or absent in
the 1-0.  Although the prominence of the shoulders is a
straightforward consequence of scattering in a semi-infinite
atmosphere, shoulders have not been reported in the observed spectra
of accretion-powered pulsars.  {\it If the shoulders persist after the
model spectra have been folded through the detector response matrix,
and they prevent an acceptable fit to an observed spectrum, their
absence in the observed spectrum could be a significant clue to the
geometry of the line-forming region\/} --- it indicates either a
geometry analogous to a 1-0 slab with no restriction on the column
depth or a geometry analogous to a 1-1 slab with a column depth
$N_{e,21} \gtrsim 15,000\ (\tau_{To} \gtrsim 10)$.

\medskip
Using both 1-1 and cylindrical geometries, \cite{ah96} consider the
effects of line shoulders on the spectrum of the accretion-powered
pulsar A0535+26.  Kendziorra {\it et al.} (1992, 1994) report HEXE
observations of 50 and 100 keV features from this pulsar at 2 and 4.5
$\sigma$ respectively.  OSSE observations by \cite{grove95} of a
feature at 110 keV confirm the high energy feature reported by HEXE;
the low energy feature is too close to the 45 keV OSSE threshold for a
conclusive observation.  \cite{ah96} consider the possibility that the
110 keV feature is a second harmonic line in a spectrum where the first
harmonic has been filled in by the line shoulders.  They fit this model
to the observed spectrum, for $N_{e,21} \lesssim 5$; the fit is poor
compared with a model in which the 110 keV feature is the fundamental.
The latter model is plausible in light of the low significance of the
50 keV feature in the HEXE observation.

\medskip
Although the column depths in Araya and Harding's (1996) calculation
are suitable for scattering in the magnetosphere (see below), they are
much smaller than the column depths usually considered typical of
accretion columns.  The results of section \ref{secSpectra} in the
present work, which show that the line shoulders disappear at large
column depths, support the conclusion of \cite{ah96} that A0535+26 does
not possess a $\sim 50$ keV first harmonic line that has been filled in
by the line shoulders.  However, the spectrum of A0535+26 could possess
a $\sim 50$ keV first harmonic that has been mostly filled in with
photons spawned by scattering at higher harmonics in a medium that is
optically thick in the wings.  It is difficult to assess this
conjecture until second and higher harmonic scattering at such optical
depths are better understood.  In addition, the fact that the first
harmonics observed in other accretion-powered pulsars are broad but
shallow, with $W_E$ much smaller than those given by theoretical
models, gives pause.

\medskip
In many accretion-powered pulsars, the pulse profiles are complex and
strongly energy-dependent, and elude explanation by simple models.
Dermer and Sturner (1991) and Sturner and Dermer (1994) suggest that
some of these properties can be explained by photon scattering in the
pulsar magnetosphere, in a layer of plasma supported by radiation
pressure.  Such an atmosphere has a geometry similar to the 1-0.  By
equating the momentum per second carried by the radiation to the
gravitational force on the suspended plasma, Sturner and Dermer
calculate the {\it maximum} mass that can be suspended at a given
altitude.  At the maximum mass, these layers are optically thick to
line scattering but thin to continuum scattering.  They are located at
a distance of a few stellar radii from the star's center.  A
calculation of the {\it actual} suspended mass needs to take into
account the scattering cross section, the effects of multiple
scatterings, and the stability of the suspended layer.  We will address
these issues in a separate paper.

\medskip
Sturner and Dermer calculate the effect of scattering in the suspended
plasma on the radiation {\it pulse profiles}.  They report that the
scattering flattens the pulse profiles.  The flattening corresponds to
the reduction in flux at the line center that we see in our Monte Carlo
spectra.  In the present work we speculate about the signature of
magnetospheric scattering on the {\it spectrum}.  If a suspended layer
could be formed, spectral features formed there would have small or no
shoulders, in agreement with observations.  At $\Psi=0$, it would be
difficult for a layer with $N_{e,21}\sim1$ to form the broad lines
required for the case of accretion-powered pulsars.  However, as the
present work shows, broad lines are not a problem as $\Psi$ approaches
$\pi/2$.  The lines would be broadened even more if the line-forming
region covers an extended region in which the magnetic field strength
varies significantly.

\medskip
However, if the lines from an accretion-powered pulsar are formed in
this way, one expects that the dipole field $B_o$ at the neutron star
surface would be $\sim 10$ times larger than the dipole component of
the field $B(r)$ in the line-forming region in the magnetosphere.  The
field $B(r)$ can be inferred from the cyclotron line energy.  For most
sources, such as Her X-1, the field inferred from the cyclotron lines
is {\it larger} than the dipole field inferred from the accretion
torque model of Ghosh and Lamb (1979a,b; 1991), contrary to what would
be expected if the lines are formed in the magnetosphere.  However, it
is possible that the {\it non-}dipole components of the field in the
line-forming region is large enough to account for the discrepancy.
There are also some accretion-powered pulsars in which the field
inferred from the cyclotron line energy could conceivably be much {\it
larger} than the dipole surface field inferred from the accretion
torque theory (Ghosh and Lamb 1979a,b, 1991; Mihara 1995).

\subsection{Gamma-ray Bursters}
\label{ssecgrb}

\medskip
The fit to the spectrum of gamma-ray burst GB880205 made by Wang {\it
et al.} (1989) assumes a static line-forming region with a uniform
magnetic field parallel to the slab normal.  As we mention in Section
\ref{intro}, this geometry is suitable, for example, for the magnetic
polar cap of a neutron star with a dipole field.  LWW point out that if
the line-forming region is indeed at the polar cap, the static model is
valid only if the bursters lie at distances less than several hundred
parsecs.  Otherwise, the burst luminosity exceeds the Eddington
luminosity and the radiation force creates a relativistic plasma
outflow along the field lines.

\medskip
However, in order to explain the brightness and sky distributions of
the burst observed by BATSE, it has been suggested that, if the
bursters are galactic, the sources are in a galactic corona at
distances of 100-400 kpc (for a review, see Lamb 1995).
In light of the BATSE results, it is important to explore
line-formation models that are appropriate for sources at these
distances.  

\medskip
One possibility is line-formation in a relativistic outflow (Miller
{\it et al.} 1991, 1992; Isenberg, Lamb, and Wang 1996).  Another
possibility is line-formation in a static slab at the magnetic
equator.  Here the field lines are parallel to the slab ($\Psi=\pi/2$),
and can confine the plasma magnetically (see, e.g., Zheleznyakov and
Serber 1994, 1995).  Using the revised Monte Carlo code developed in
the present work, and varying $\Psi$, Freeman {\it et al.} (1996) fit
models to the two observed spectra corresponding to the two time
intervals S1 and S2 in which lines appeared in burst GB870303.  When
they perform a joint fit to the two intervals, using models with a
common $B_{12}$ and $N_{e,21}$ for both spectra but not a common
viewing angle, they find that in the 1-1 geometry the data marginally
favors the equatorial model over the polar model.

\medskip
One promising line of future research is to calculate spectra for a
region of a neutron star's surface large enough so that the field
strength and orientation vary over the region.  By comparing such
calculations to observations, we can put a limit on how large a
radiating area is compatible with the narrow lines found in gamma-ray
bursts.

\medskip
There are still many open issues concerning both models.  For example,
for the large magnetic fields and source distances required by the
galactic corona model, the optical depths for the processes $\gamma
\rightarrow e^+ e^-$ and $\gamma \gamma \rightarrow e^+ e^-$ are both
much larger than unity (Schmidt 1978; Daugherty and Harding 1983; Burns
and Harding 1984; Brainerd and Lamb 1987).  The production of
electron-positron pairs by these processes could lead to line-forming
regions that are optically thick in the continuum, which might prevent
the formation of the narrow lines observed in gamma-ray bursts.  In
addition, pair production could truncate the spectrum at the pair
production threshold, 1 MeV.  Truncation is inconsistent with
observations by COMPTEL (Winkler {\it et al.} 1993) and EGRET (Schneid
{\it et al.} 1992; Kwok {\it et al.} 1993; Sommer {\it et al.} 1994) of
photon energies up to 1 GeV with no evidence of a spectral cutoff or
rollover at $\approx 1$ MeV.

\medskip
Various solutions to these problems have been proposed (for a review,
see Higdon and Lingenfelter 1990).  One possibility is that the entire
spectrum is produced at the magnetic polar cap in a medium with bulk
motion corresponding to a Lorentz factor of $\Gamma \sim 10$.  The
radiation would then be beamed into an angle $\sim 1/\Gamma$.  For
corona distances, the optical depth for the two photon process would
then be reduced enough to be consistent with the observed spectra above
1 MeV (Schmidt 1978; Baring and Harding 1993; Harding and Baring 1994;
Harding 1994). Additional calculations are necessary to determine
whether the observed cyclotron features could be formed at the polar
cap under these physical conditions.

\medskip
Alternatively, the gamma-ray burst spectrum could consist of two
components produced in two distinct physical regions: a soft ($\lesssim
1$ MeV) component produced in plasma trapped in the equatorial region
of the magnetosphere (Lamb 1982; Katz 1982, 1994) and a hard component
($\gtrsim 1$ MeV) produced by relativistic outflow at the magnetic
polar cap.  Because the magnetic field traps the plasma in the
line-forming region, the line-forming region can be static, even for
highly super-Eddington luminosities, as is the case if the bursts come
from neutron stars in a galactic corona.  The hard component could be
produced in a relativistic pair outflow or expanding fireball at the
magnetic polar cap.  If so, the cross section for the two photon
process is further reduced by the bulk motion of the pairs.  A recent
report by \cite{cm95} of evidence for two components in the spectrum of
GB881024 is intriguing.  The soft component dominates the spectrum for
$E<250$ keV; the hard component is dominant above this energy.

\acknowledgments  
We wish to thank Tomasz Bulik, Deepto Chakrabarty, Peter Freeman, Cole
Miller, and Rob Nelson for helpful discussions.  This work was
supported in part by NASA Grants NAGW-830, NAGW-1284, and NAG5-2868 at
Chicago, and NASA Grant NAG5-3836 at Maryland.  MI gratefully
acknowledges the support of a NASA Graduate Student Researchers
Fellowship under NASA Grant NGT-5-18.

\appendix
\section{Appendix -- Calculation of ${\bf T_c/E_B}$ in Optically Thin Slabs}
\label{secapp}

Start with eq. (\ref{Dele}):

\begin{equation}
\Delta E \equiv {{\int\,d\Omega\,dE\, n(E,\Omega)\, \,
    {\overline{\delta E}}\,N_{scat}(\tau_{To},E,\Omega)}\over
    {\int\,d\Omega\,dE\,n(E,\Omega)}}\ ,. \eqnum{\ref{Dele}}
\end{equation} 

\noindent
 In terms of the differential scattering cross section
 $d\sigma/d\Omega_s$,

\begin{eqnarray}
{\overline{\delta E}} & \equiv & {{\int\,dp\, f(p)
  \,d\Omega_s\,{{d\sigma }\over {d\Omega_s }}\,\delta E }\over
  {\int\,dp\, f(p) \,d\Omega_s\,{{d\sigma }\over {d\Omega_s }} }}
  \nonumber \\ & = & {\int\,dp\,d\Omega_s\, f(p) \,{{1 }\over
  {\sigma_T \phi(E,\Omega) }}{{d\sigma }\over{d\Omega_s }}\,\delta E}\ ,
\label{Mean}
\end{eqnarray} 

\noindent
where $\delta E=E_s-E$ is the energy transferred from the electron to the
photon in an individual scatter and $\phi$ is as defined in equation
(\ref{ScatProfile}).   Substituting
    eq. (\ref{Mean}) into (\ref{Dele}) and adopting a normalized photon
    distribution, we obtain

\begin{equation}
\Delta E = \int\,d\Omega\,dE\,dp\,d\Omega_s\, n(E,\Omega)\, f(p) \,
    {{1 }\over {\sigma_T \phi}}{{d\sigma }\over{d\Omega_s }}\,\delta
    E\, N_{scat}(\tau_{To},E,\Omega)\ .
\label{Delep}
\end{equation} 

   In the optically thin,
   or single scattering limit, the total number of scatters per photon
   is just the fraction that experienced scattering, viz,

\begin{equation}
N_{scat}(\tau_{To},E,\Omega)=1-\exp\left[-{{\tau(E,\Omega) }\over
     {\vert\mu_{sl}\vert }}\right]\approx {{\tau }\over{\vert\mu_{sl}\vert
     }}\ ,
\label{nscat}
\end{equation} 

\noindent
  where for an isothermal medium 

\begin{equation}
{\tau \over \vert\mu_{sl}\vert} =
     {\tau_{To}\phi(E,\Omega) \over \vert\mu_{sl}\vert} \ \ \ll 1\ ,
\end{equation}

\noindent
is the depth at which photons are injected measured from the top of
the slab, along the line of sight.
Substituting eq. (\ref{nscat}) into (\ref{Delep}) gives

\begin{eqnarray}
\Delta E & = 
   N_e\,\int_{-\infty}^{+\infty}\,dp\, f(p) \,
   \int\,{{d\Omega}\over{\vert\mu_{sl}\vert }}\, dE\,n(E,\Omega)\,
    \left[ \int\,d\Omega_s\, {{d\sigma }\over{d\Omega_s }}\,E_s -
    E \int d\Omega_s\,{d\sigma \over d\Omega_s}\ \right ]\ ,
\label{Delef}
\end{eqnarray} 

\noindent
 where we have used $\delta E = E_s - E$.

\medskip
 For first harmonic resonant scattering with zero natural line width,
 the differential cross section is given by eq. (\ref{ResCross})
in the limit $\Gamma_1 \rightarrow 0$
 ($\hbar=c=k=1$ used throughout):

\begin{equation}
{{d\sigma^r}\over{d\Omega_s^r }} = {9\over
       32}\,{\sigma_T m_e \over \alpha}\,\delta(E^r-E_1^r)\,
       {1+{\mu^r}^2 \over 2}\,{1+{\mu_s^r}^2 \over 2}\ .
\label{xsec}
\end{equation} 

\medskip
   For atmospheres with $T_e\ll m_e$ and $b\ll 1$, the two naturally
   occurring small parameters are the electron velocity along the
   field [$\beta\sim (T_e/m_e)^{1/2}$] and the gyration velocity
   orthogonal to the field ($\beta_{gyr}^2\sim b \sim E/m_e $). 
   To evaluate $T_c$, we expand eq. (\ref{Delef}) to $O(\beta^2)$ and
$O(b\sim E/m_e)$.  
   Thus, the scattered frequency is given by $E_s=\gamma
   E_s^r (1+\beta\mu_s^r)$, where

\begin{equation}
 E_s^r = E^r\left\{ 1 - {{E^r }\over{2m_e
    }}\,\left(\mu_s^r-\mu^r\right)^2 + O\left[ \left( {{E^r
    }\over{2m_e }} \right)^2 \right] \right\}.
\label{escat}
\end{equation} 

\medskip
We assume that the photon density is separable so that

\begin{equation}
n(E,\Omega) = n(E)\,Q(\Omega).
\label{sep}
\end{equation} 

\medskip
Substituting eqs. (\ref{siglab}) and (\ref{xsec})--(\ref{sep}) into (\ref{Delef}), integrating over
scattered angles, and integrating over $E$ using

\begin{equation}
  \delta(E^r - E_1^r) = {1\over{\gamma(1-\beta\mu)}}\,\delta(E-E_1)\ ,
\label{deltalab}
\end{equation} 

\noindent
   where $E_1 = E_1^r/[\gamma(1-\beta\mu)]$, gives

\begin{eqnarray}
\Delta E \approx {3 \tau_{To} m_e \over 16 \pi \alpha} N_e\,
\int_{-\infty}^{+\infty}dp\,f(p)\, 
\int\,{{d\Omega }\over{\vert\mu_{sl}\vert}}\,Q(\Omega)\,N(E_1)\,
E_1^r\,(1+{\mu^r}^2)\, \times\nonumber \\ \left[ 1 -
{1\over{\gamma^2(1-\beta\mu)}} - {{E_1^r }\over{m_e }}\,\left(
{{{\mu^r}^2 }\over{2 }} + {1\over 5}\right) \right].
\label{Delefp}
\end{eqnarray} 

\noindent
   The leading order term in brackets is $O(\beta)$ and so we need
   only expand $N(E_1)E_1^r(1+{\mu^r}^2)$ to $O(\beta)$ to obtain the
   desired result. Doing this gives

\begin{eqnarray}
\lefteqn{\Delta E = {3 \tau_{To} m_e\over 16 \pi \alpha}\,N(E_B)E_B\,
       \int_{-\infty}^{+\infty}dp\,f(p)\, \int\,{{d\Omega
       }\over{\vert\mu_{sl}\vert}}\,Q(\Omega)\, \times} \nonumber \\ &
       \qquad \bigg\{ -\beta\mu(1+\mu^2) +
       \beta^2[1+(s+2)\mu^2+(s-3)\mu^4] - b(1+\mu^2)\left({{\mu^2
       }\over{2}}+{1\over 5}\right) \nonumber \\ & 
         + O(\beta^3,b\beta) \bigg\}\ ,
\label{penD}
\end{eqnarray} 

\noindent
where

\begin{equation}
s \equiv -{E\over{n(E)}}\,{{dn }\over{dE }}\Bigg\vert_{E=E_B}.
\label{indx}
\end{equation}

\noindent
Note that if the initial photon spectrum is a power law, $n(E) \propto
E^{-\alpha}$, then $s$ is just the spectral index, $\alpha$.

\medskip
We assume $f(p)$ to be a non-relativistic one-dimensional Maxwellian, 
that is,

\begin{equation}
f(p)\,dp = {\exp\left(-p^2 \over 2 m_e T_e \right)\over \sqrt{2\pi m_e T_e }}\,dp\ .
\label{max}
\end{equation} 

\noindent
   Integrating over $p$ in eq. (\ref{penD})
     and setting $\Delta E=0$ gives
      the slab Compton temperature

\begin{equation}
{T_c \over {E_B}} = {1\over{10}}\,{{ \int{{d\Omega
     }\over{\vert\mu_{sl}\vert}}\,Q(\Omega)\, (2+7\mu^2+5\mu^4)}\over{
     \int{{d\Omega }\over{\vert\mu_{sl}\vert}}\,Q(\Omega)\,
     [1+(s+2)\mu^2+(s-3)\mu^4] }}\ .
\label{tceq}
\end{equation}

\begin{figure}
\figcaption{Line-forming region geometries.  (a) Slab geometry with
magnetic field direction at an angle $\Psi$ to the slab normal.  The
slab is infinite in extent.  $N_e$ is the column depth between the
source plane and the top surface; $N_e^\prime$ is the column depth
between the bottom surface and the source plane.  $\theta_{sl}$ and
$\varphi_{sl}$ denote the polar and azimuthal angles of the photon
direction with respect to the slab normal respectively.  (b)
Cylindrical geometry with magnetic field parallel to cylinder axis.
The cylinder is infinite in length.  $N_e$ is the column depth between
the photon source, located along the cylinder axis, and the surface.
In both geometries, $\theta$ is the angle between the photon direction
and the magnetic field direction.
\label{geometry}}
\end{figure}

\begin{figure}
\figcaption{ Scattering profiles vs. photon energy for field strength
$B_{12}=3.5$ ($E_B=40$ keV), $T_e=10$ keV, and incident photon
direction cosine $\mu=$ 0, 0.5, and 1.
Panels (a)--(c) $\phi_0$ ({\it dots}), $\phi_1$ ({\it dashes}), and
$\phi_0+\phi_1$ ({\it solid}), including non-Lorentzian factor $4E^{r3}
E_B^{-1} (E^r + E_B)^{-2}$.  Panels (d)--(f) Comparison of $\phi_0+\phi_1$ with
({\it solid}) and without ({\it dot-dash}) non-Lorentzian factor.
\label{profile}}
\end{figure}

\begin{figure}
\figcaption{Resonant Compton temperature $T_c/E_B$
as a function of the angle
$\Psi$ between the magnetic field and the slab normal for optically thin 
models ($\tau / \vert \mu_{sl} \vert << 1$) with
$B_{12}$ = 1.7, first harmonic scattering only, and zero natural line
width.  The {\it solid curve} is the analytic result 
with non-relativistic scattering kinematics.
Monte Carlo calculations using the 1-1 line-forming region geometry
are shown for non-relativistic kinematics ({\it squares}), relativistic kinematics with both $p_{max}$ and
$p_{min}$ channels open ({\it triangles}), and relativistic kinematics
with only the $p_{min}$ channel open ({\it circles}).
\label{tcthin}}
\end{figure}

\begin{figure}
\figcaption{Monte Carlo calculations of the equilibrium
Compton temperature $T_c/E_B$ as a function of the angle $\Psi$
between the magnetic  
field and
the slab normal for slabs that are optically thick in the
line core. The calculations use $N_{e,21} = 1.2$ and $B_{12}$=1.7.  
Temperatures are shown
for line-forming regions with both the 1-0 ({\it circles}) and 1-1 ({\it
triangles}) geometries.
\label{tcthick}}
\end{figure}

\begin{figure}
\figcaption{ The mean energy associated with the net photon flux as a function of
 depth inside a slab with field strength $B_{12}=3.5$, field
 orientation $\Psi=0$, and column depth $N_{e,21}=15,000\
 (\tau_{To}=10)$.  The source plane is at $\tau_T/\tau_{To}=1$ and the
 upper surface is at $\tau_T/\tau_{To}=0$.  Results for the 1-1 ({\it
 solid}) and 1-0 ({\it dashes}) geometries are shown.  The mean energy
 at the source plane in the 1-1 geometry is zero because of the net
 zero flux at this location.
\label{netflux}}
\end{figure}

\begin{figure}
\figcaption{Cumulative injected ({\it dots}) and emerging ({\it solid})
spectra for a slab with field strength $B_{12}=3.5$, field orientation
$\Psi=0$, and column depth $N_{e,21}=15,000\ (\tau_{To}=10)$.  The
injected spectra is a 10 keV Wien distribution.  Even though only 14\%
of the photons are injected with $E<E_{\rm thin}(7)=12.9$ keV ({\it
dashes}), 45\% escape below this energy.  The glitch in the emerging
spectrum just above $E_{\rm thin}(7)$ is due to the random
fluctuations in the Monte Carlo simulation.  The output bin for this
energy holds an unusually low number of photons.
\label{cumspec}}
\end{figure}

\begin{figure}
\figcaption{ Spectra for $\Psi=0$ and several viewing angles.
$B_{12}$=1.7 and $N_{e,21}$=1.2.  Monte Carlo spectra for the 
1-1 ({\it solid}) and 1-0 ({\it dots}) geometries are shown, as well as 
relativistic absorption spectra with finite natural line width ({\it
dashes}).
\label{spectra0}}
\end{figure}

\begin{figure}
\figcaption{ Spectra for $\Psi=\pi/4$ and several viewing angles.
$B_{12}$=1.7 and $N_{e,21}$=1.2.  Monte Carlo spectra 
for the 1-1 ({\it solid}) and 1-0 ({\it dots}) geometries are shown,
as well as 
relativistic absorption spectra with finite natural line width ({\it
dashes}).
\label{spectra45}}
\end{figure}

\begin{figure}
\figcaption{ Spectra for $\Psi=\pi/2$ and several viewing angles.
$B_{12}$=1.7 and $N_{e,21}$=1.2.  Monte Carlo spectra 
for the 1-1 ({\it solid}) and 1-0 ({\it dots}) geometries are shown,
as well as
relativistic absorption spectra with finite natural line width ({\it
dashes}).
\label{spectra90}}
\end{figure}

\begin{figure}
\epsscale{0.95}
\figcaption{ Equivalent width of the first harmonic line as a function of
viewing angle in the 1-0 geometry.  The calculations use $B_{12}$=1.7,
$N_{e,21}$=1.2, 
$T_c$=8.38 keV, and $\Psi=\pi/4$. 
The equivalent widths become
negative (i.e., there are emission-like features) as $\mu_{sl}$
approaches unity.
The surface is symmetric about $\varphi_{sl}=\pi/2$.
\label{ewa}}
\end{figure}

\begin{figure}
\epsscale{0.95}
\figcaption{ Equivalent width of the first harmonic line as a function of
viewing angle in the 1-1 geometry.  The calculations use $B_{12}$=1.7,
$N_{e,21}$=1.2, 
$T_c$=9.76 keV, and $\Psi=\pi/2$.  
The equivalent widths become
negative (i.e., there are emission-like features) as $\mu_{sl}$
approaches unity.
The surface is symmetric about $\varphi_{sl}=\pi/2$.
\label{ewb}}
\end{figure}

\begin{figure}
\epsscale{1}
\figcaption{First harmonic equivalent width as a function of
$\mu_{sl}$ for the 1-0 ({\it dashes}) and 1-1 ({\it solid})
geometries for different values of $\Psi$ and $N_{e,21}$.  
$B_{12}$=1.7.
\label{ewc}}
\end{figure}

\begin{figure}
\figcaption{Angular distribution of emerging radiation for isotropic
photon injection at $E=E_B$,  $B_{12}=1.7$ and $N_{e,21}=1.2$.
The spectrum $N(\mu_{sl})$ is summed over energies and averaged over
the azimuthal angle $\varphi_{sl}$.
In the 1-0 geometry, the angular distributions of both the transmitted ({\it solid}) and reflected 
({\it dashes}) spectra are shown.  In the 1-1 geometry the transmitted
and reflected spectra are identical, by symmetry.  The distributions
are normalized 
so that the sum of the transmitted and reflected spectra is equal to
unity.  
The photons tend to escape transverse
to the slab, regardless of the direction of the magnetic field.
\label{binhist}}
\end{figure}

\begin{figure}
\epsscale{1}
\figcaption{ Shoulder formation for $\Psi=0$.  Spectra for absorption
({\it dashes}) and scattering ({\it dots}) from the first harmonic
only are shown.  When higher harmonics (and spawning) are included
({\it solid}), the shoulders are enhanced.
\label{shoulders0}}
\end{figure}

\begin{figure}
\figcaption{ Shoulder formation for $\Psi=\pi/2$.  Spectra for absorption
({\it dashes}) and scattering ({\it dots}) from the first harmonic
only are shown.  When higher harmonics (and spawning) are included
({\it solid}), the shoulders are enhanced.  \label{shoulders90}}
\end{figure}

\begin{figure}
\figcaption{ Angle-integrated energy spectra for 1-1 atmospheres with
$B_{12}=1.7$, $N_{e,21}=1.2$, 
$T_e=T_c-2$ keV (a,b), $T_c$ (c,d), and $T_c+2$ (e,f),
and for both $\Psi=0$ and $\Psi=\pi/2$.
 For $T_e<T_c$,
the majority of the scattered power escapes the 
atmosphere in the red line shoulder.  For $T_e>T_c$, the majority of the
scattered power escapes in the blue line shoulder.  
\label{shoulders_vs_t}}
\end{figure}

\begin{figure}
\figcaption{ Angle-integrated flux of photons with $\mu_{sl}>0$ at 
four points inside the slab for field
strength $B_{12}=1.7$, field orientation $\Psi=0$, and column
depth $N_{e,21}=1.2$.  The flux is shown for the 1-1
({\it solid}) and 1-0 ({\it dots}) geometries.  
 The source plane is at $\tau_T/\tau_{To}=1$ and the upper surface is at 
 $\tau_T/\tau_{To}=0$.
\label{flux0}}
\end{figure}

\begin{figure}
\figcaption{ Angle-integrated flux of photons with $\mu_{sl}>0$ at several
points inside the slab for field
strength $B_{12}=3.5$, field orientation $\Psi=0$, column
depths $N_{e,21}=30\ (\tau_{To}=0.02)$, $1,500\ (\tau_{To}=1)$,
and $15,000\ (\tau_{To}=10)$, and a Wien injected spectrum with
$kT_\gamma = kT_e = 10$ keV.  The flux is shown for the 1-1
({\it solid}) and 1-0 ({\it dots}) geometries.  The cyclotron energy
is indicated by the vertical dashed line.
\label{flux}}
\end{figure}

\newpage
\begin{figure}
\figcaption{Spectra for cylindrical line-forming region with magnetic
field $B_{12}$=1.7 oriented along the cylinder axis.  The column
density from the cylinder axis to the surface is $N_{e,21}=1.2$.  The
viewing angle is characterized by $\mu=\cos \theta$ where $\theta$ is
the angle between the magnetic field and the line of sight.  Monte
Carlo spectra for scattering ({\it solid})  and absorption ({\it
dashes}) are shown.
\label{cyl}}
\end{figure}

\newpage\ 
\begin{deluxetable}{ccrr}
\tablecaption{Dependence of resonant Compton temperature on
orientation of the magnetic field, higher harmonics, 
and natural line width \label{Ttemps}} \tablehead{\colhead{Channels
open}& \colhead{Natural line width}& \colhead{$T_c/E_B\ (\Psi=0)$}&
\colhead{$T_c/E_B\ (\Psi=\pi/2)$}} 
\startdata 
\cutinhead{1-0 Geometry}
first three harmonics&
finite& $0.299^{+0.005}_{-0.004}$& $0.425^{+0.007}_{-0.008}$\nl
$0\to1\to0$ only& finite& $0.274^{+0.005}_{-0.005}$&$0.350^{+0.010}_{-0.012}$\nl 
$0\to1\to0$ only& zero&$0.262^{+0.005}_{-0.005}$& $0.352^{+0.009}_{-0.009}$\nl
\cutinhead{1-1 Geometry}
first three harmonics&
finite& $0.357^{+0.003}_{-0.003}$& $0.495^{+0.006}_{-0.006}$\nl
$0\to1\to0$ only& finite& $0.363^{+0.003}_{-0.004}$&
$0.469^{+0.008}_{-0.007}$\nl 
$0\to1\to0$ only& zero&$0.361^{+0.004}_{-0.004}$& $0.514^{+0.010}_{-0.009}$\nl \enddata
\tablecomments{Compton temperatures are for $B_{12}=1.7$ and
$N_{e,21}=1.2$. 
The effect of the higher
harmonics and natural line width is small.
 The quoted errors are statistical.}
\end{deluxetable}

\begin{deluxetable}{rcrrrrr}
\tablecaption{Large optical depth simulations.  For all simulations,
$B_{12}=3.5$, $\Psi=0$, and the injected photons have a Wien
distribution with temperature $T_\gamma=T_e=10$ keV.
\label{Tparms}} 
\tablecolumns{7}
\tablehead{
& \colhead{$0\rightarrow 0$ channel} & & & & & \colhead{$\langle E_t(\tau_T=0)\rangle$}\nl
\colhead{$N_{e,21}$}&   \colhead{open}&  \colhead{$\tau_{To}$}&   \colhead{$\tau_1$}&   \colhead{$a\tau_1$}&   \colhead{$N_{scat}$\tablenotemark{a}}&   \colhead{(keV)}\nl
}
\startdata 
\cutinhead{1-0 Geometry}
30&		Yes&	0.02&	280&	0.56&   39&   26.2\nl
1,500&		Yes&	1&	14,000&	  28&   3,280&  19.3\nl
15,000&		Yes&	10&	140,000&  280&	64,000&  14.3\nl
\cutinhead{1-1 Geometry}
30&		Yes&	0.02&	280&	0.56&   435&   28.8\nl
1,500&		Yes&	1&	14,000&	  28&   22,300&  24.6\nl
1,500& 		No&	1&	14,000&	  28&  20,800&  24.9\nl
15,000&		Yes&	10&	140,000&  280&	233,000&  15.5\nl
15,000\tablenotemark{b}&Yes&   10&     140,000&  280&  806,000&	24.7\nl
15,000& 	No&	10&	140,000&  280&  173,000&  16.4\nl
150,000&        Yes&  	100&	1,400,000&2,800&5,560,000& 10.6\nl
150,000&        No&  	100&	1,400,000&2,800&1,080,000&  9.3\nl

\enddata
\tablenotetext{a}{Number of scatterings per transmitted photon}
\tablenotetext{b}{Cross section does not include non-Lorentzian factor}
\end{deluxetable}

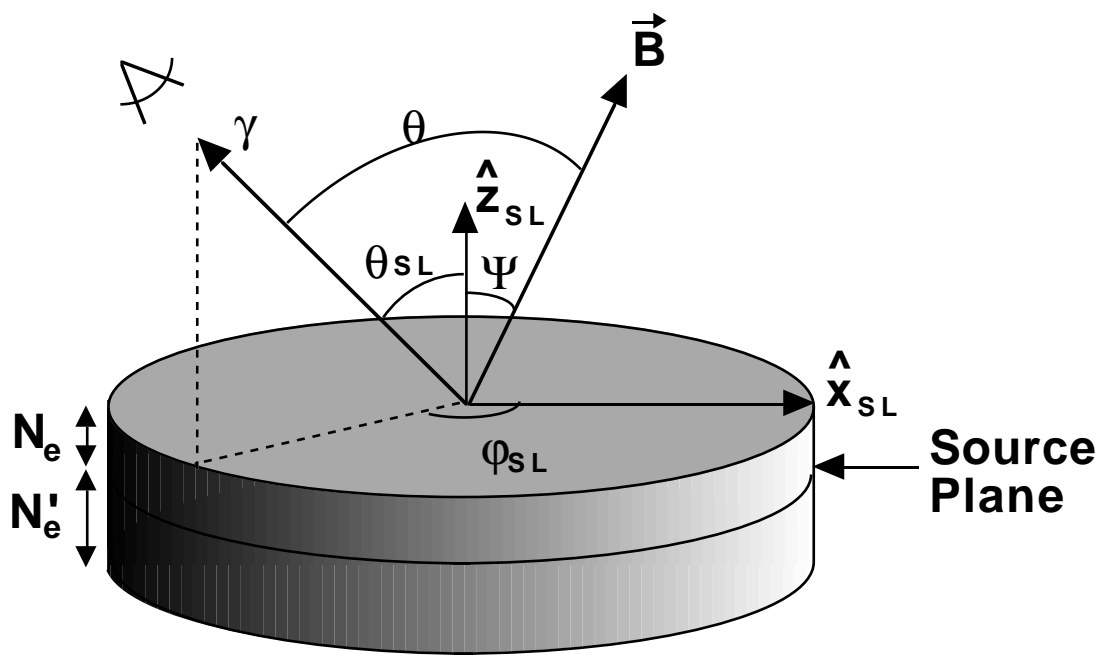
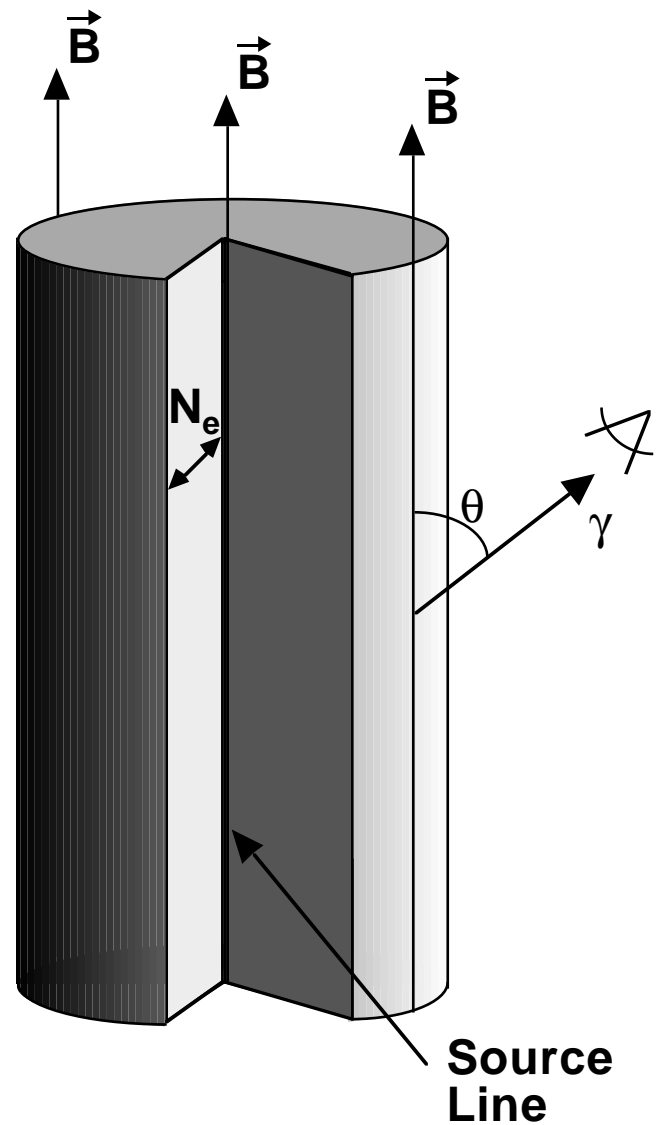
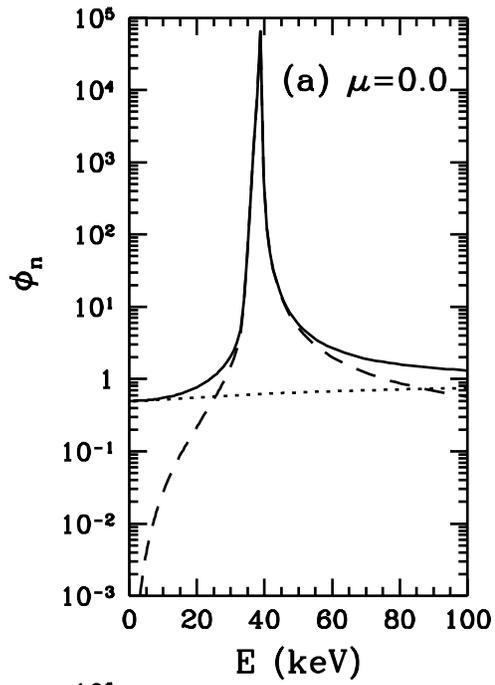
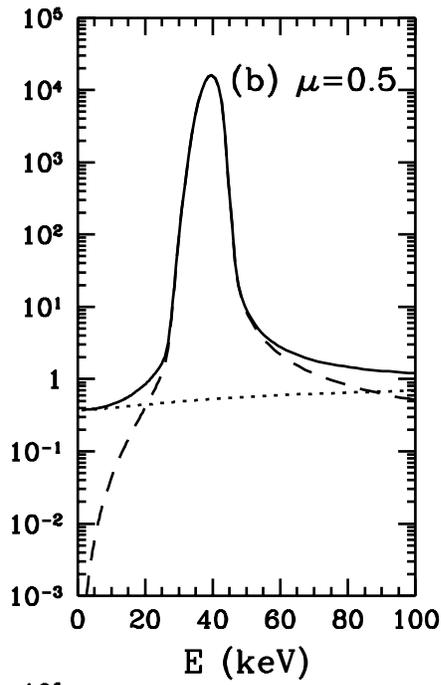
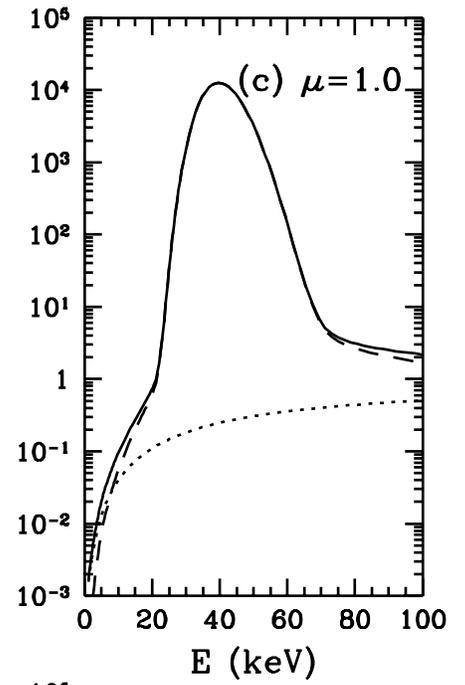
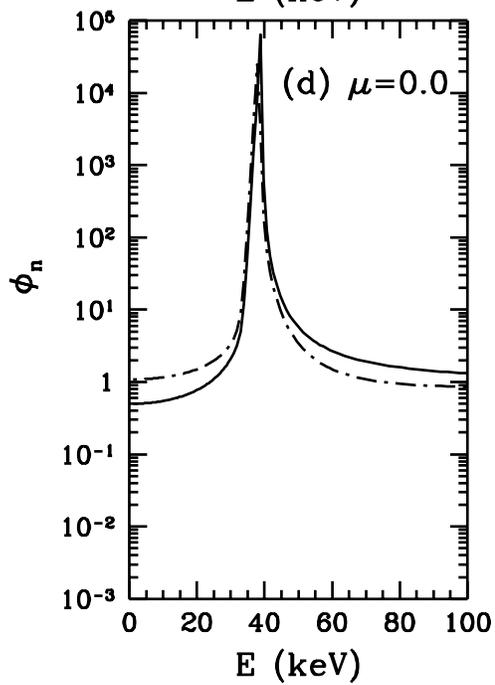
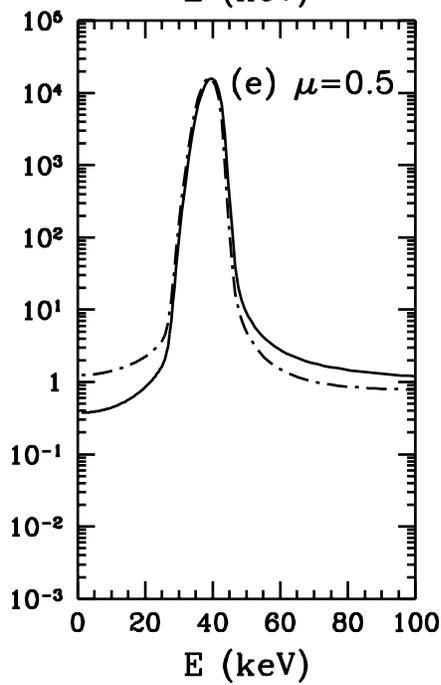
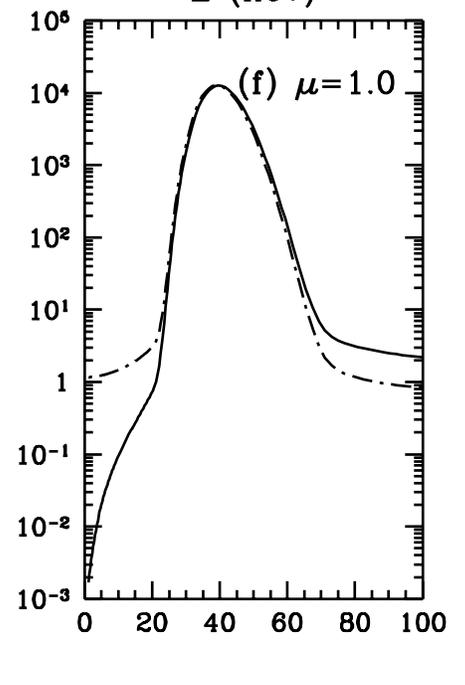
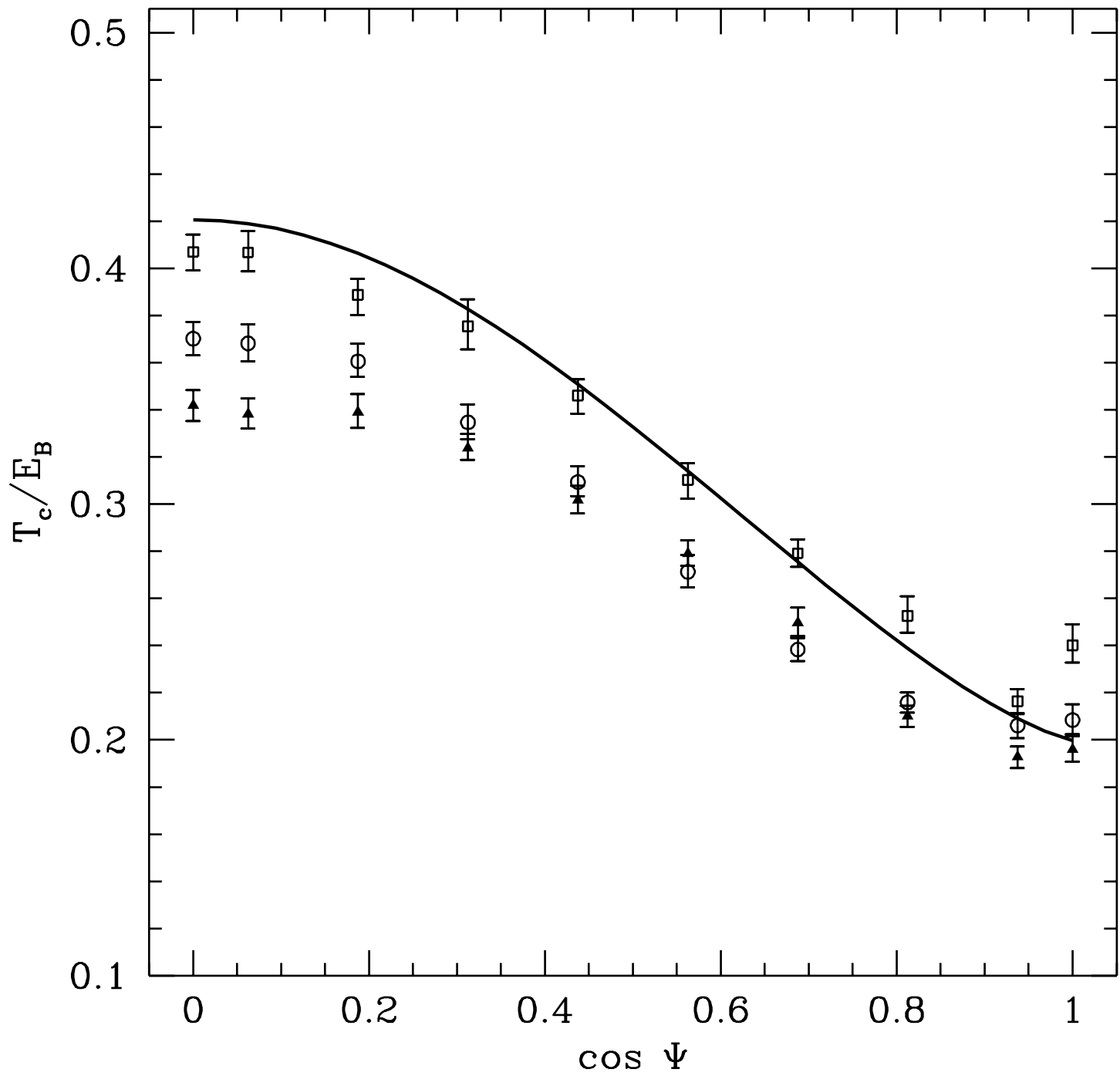
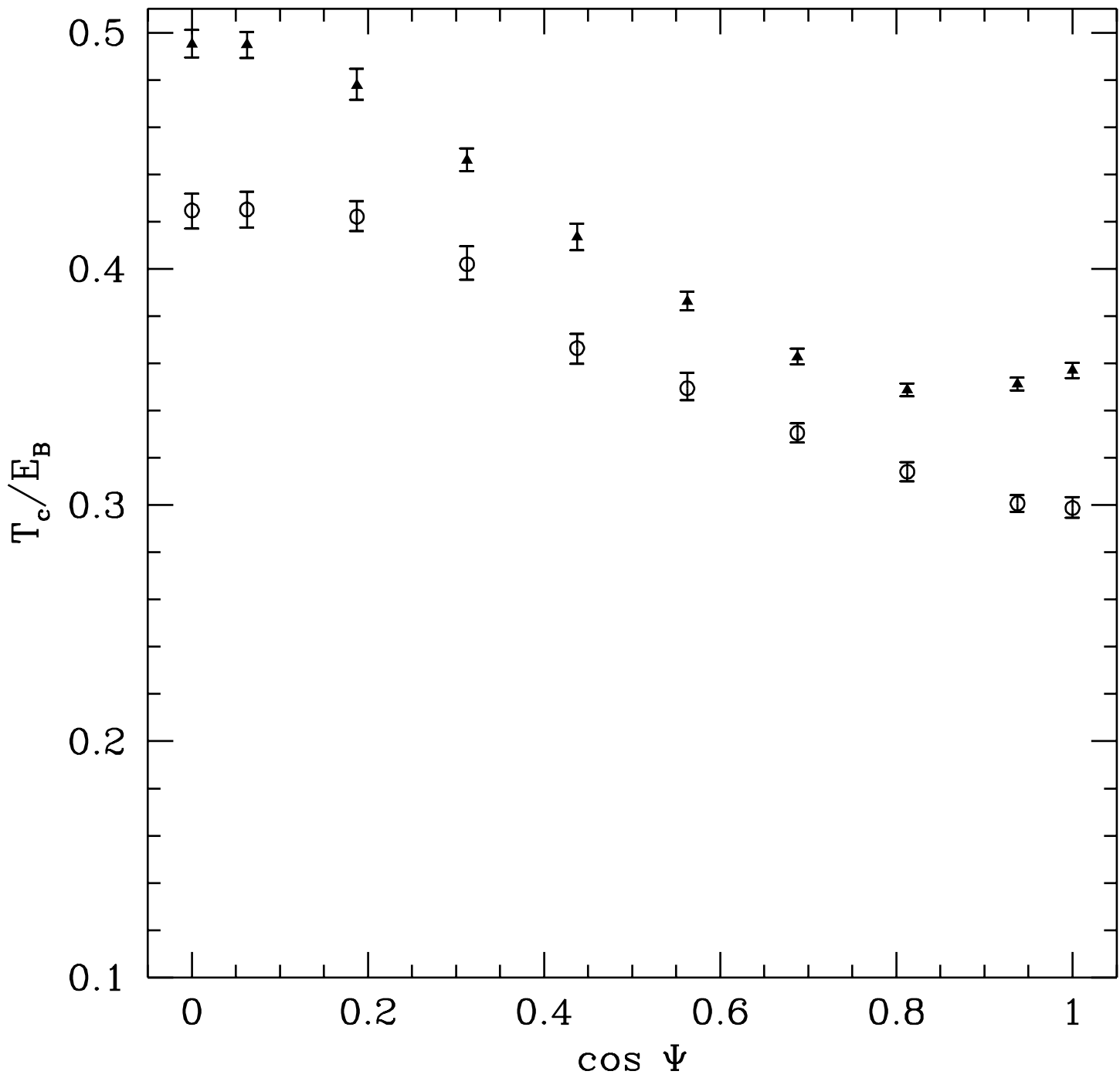
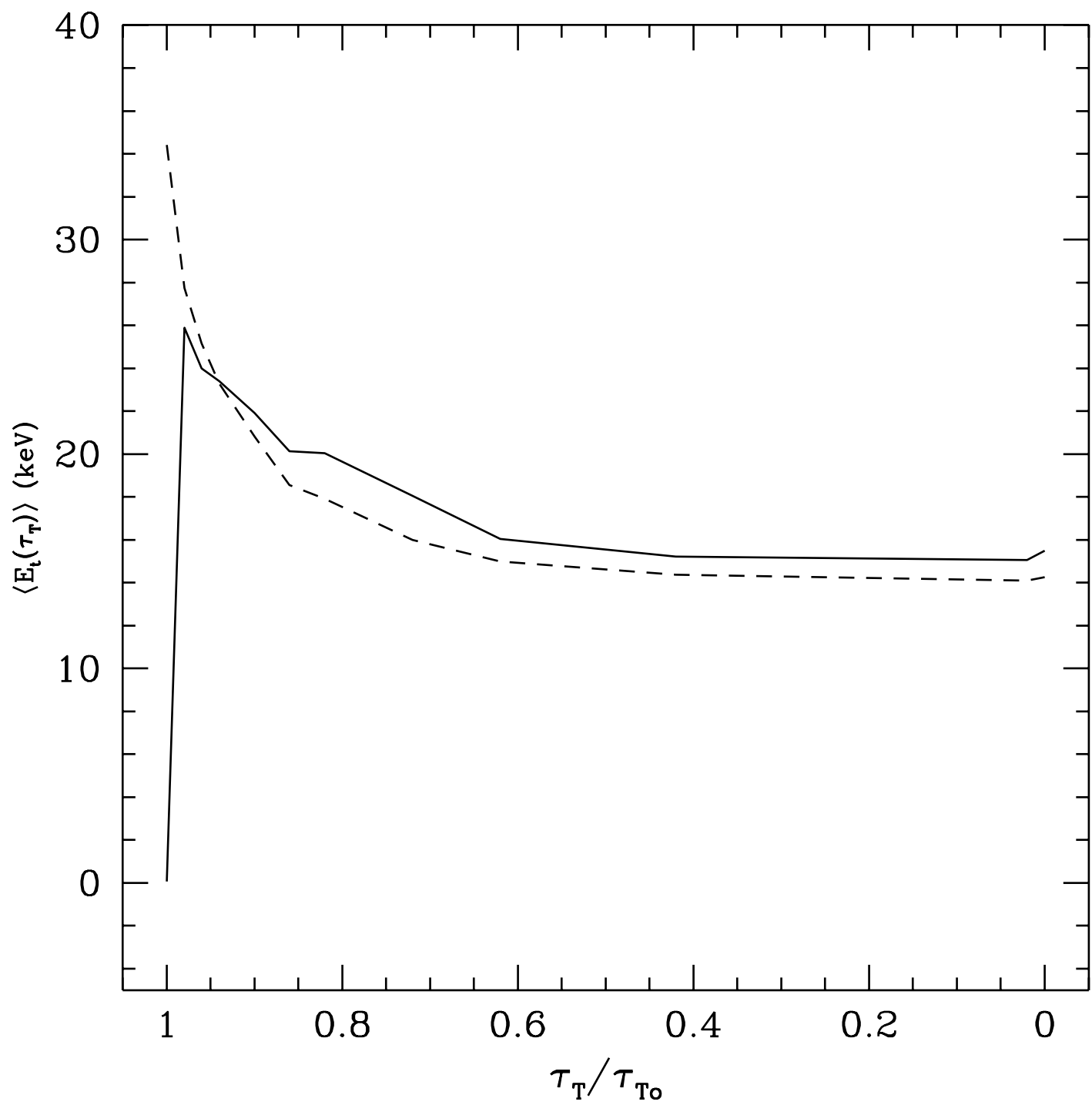
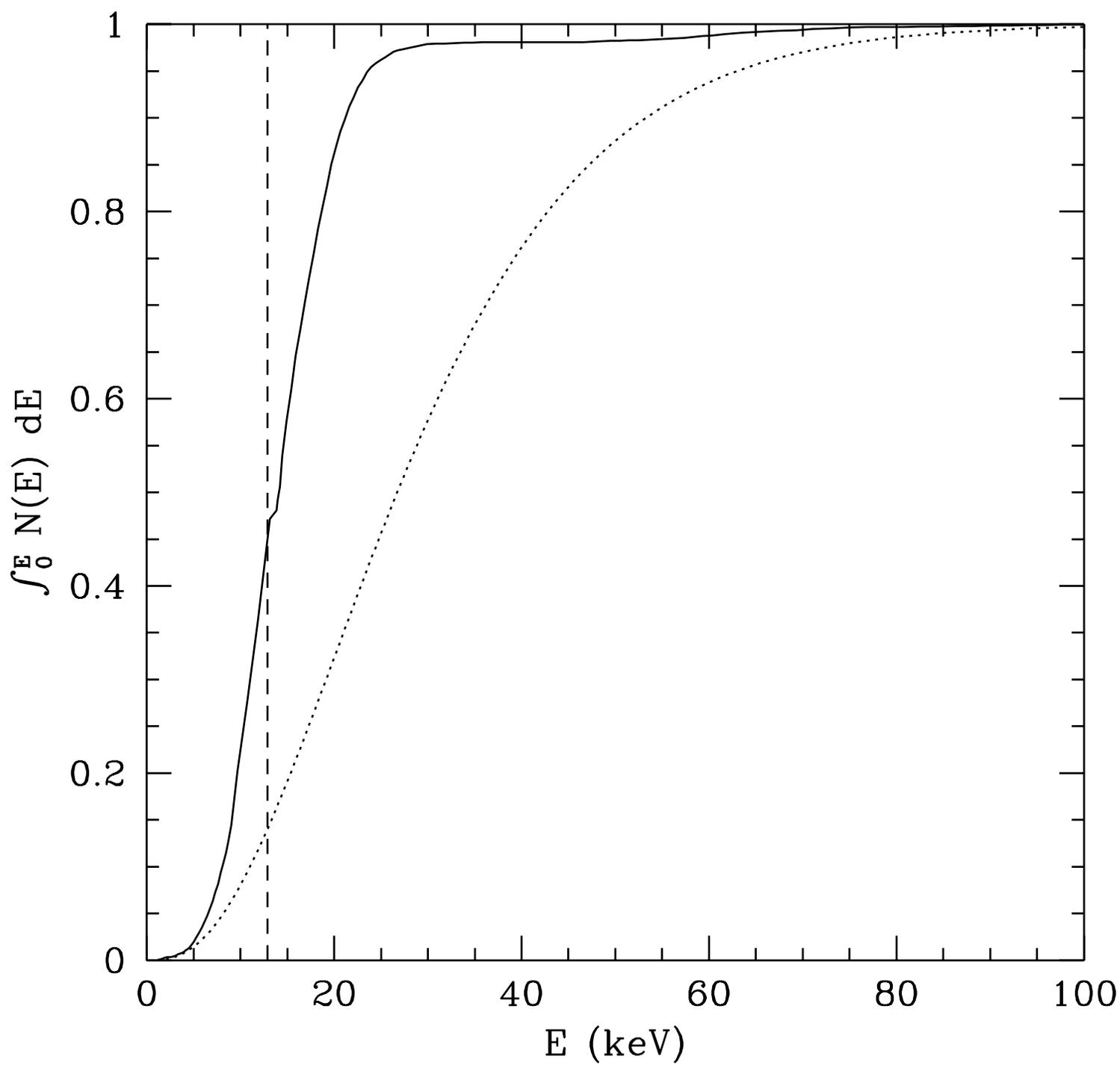
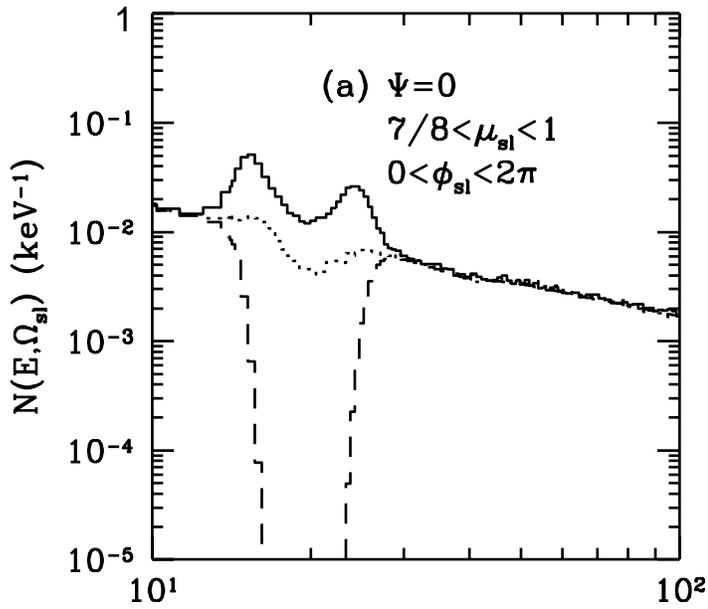
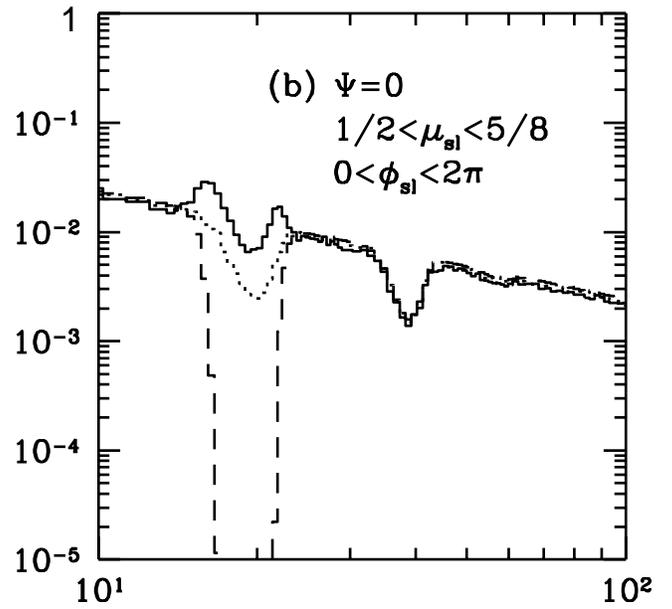
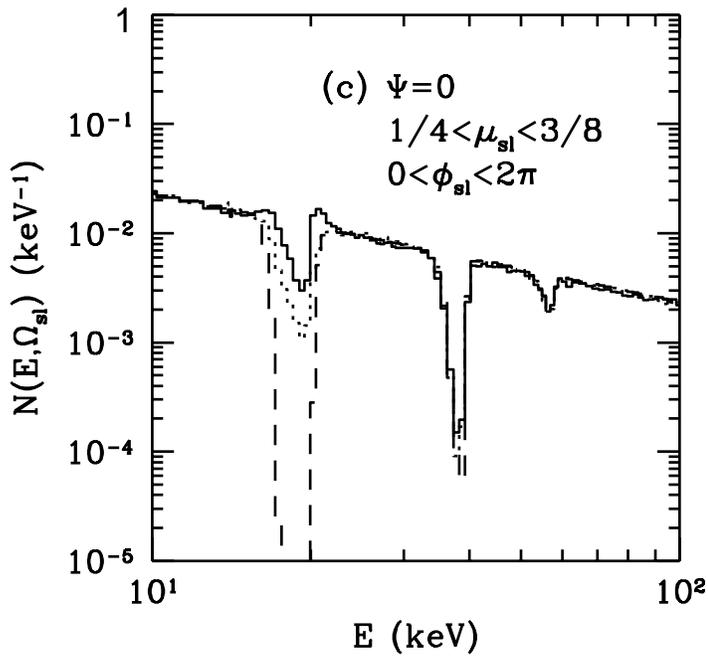
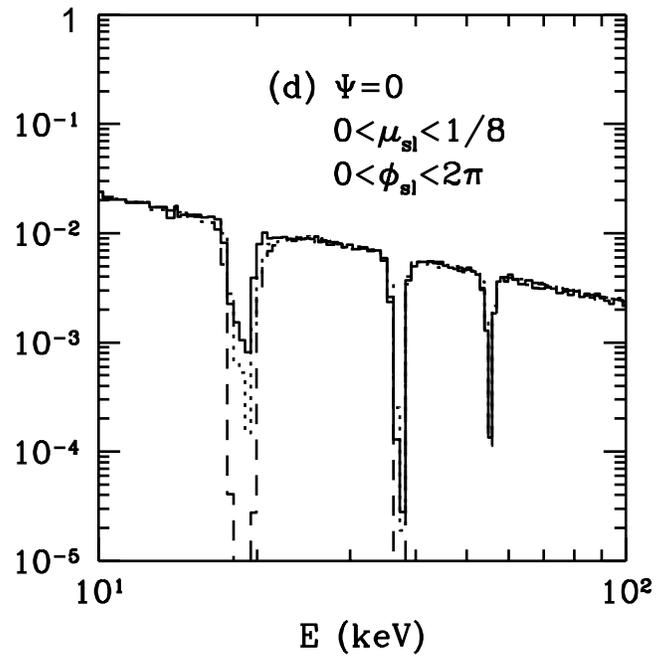
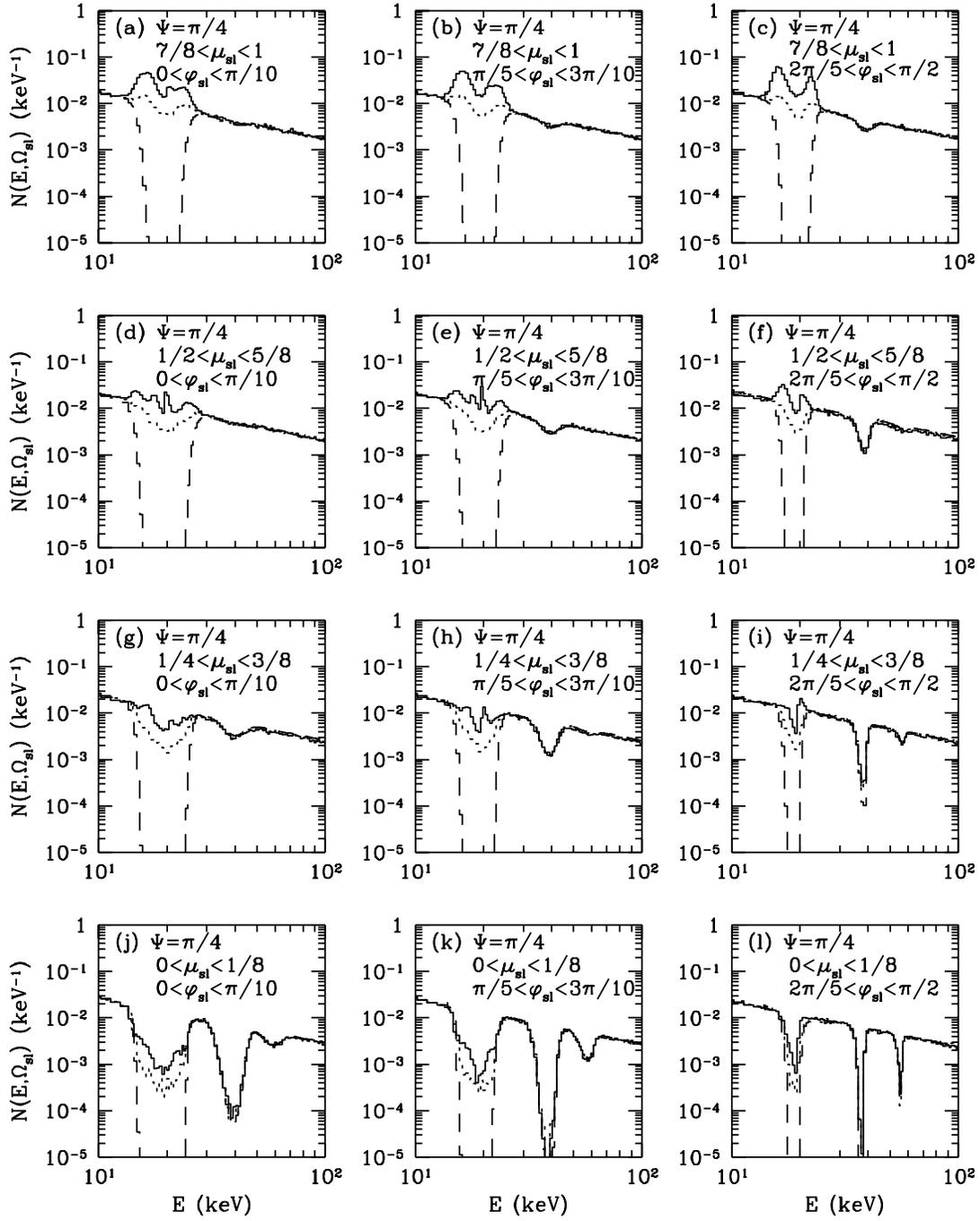
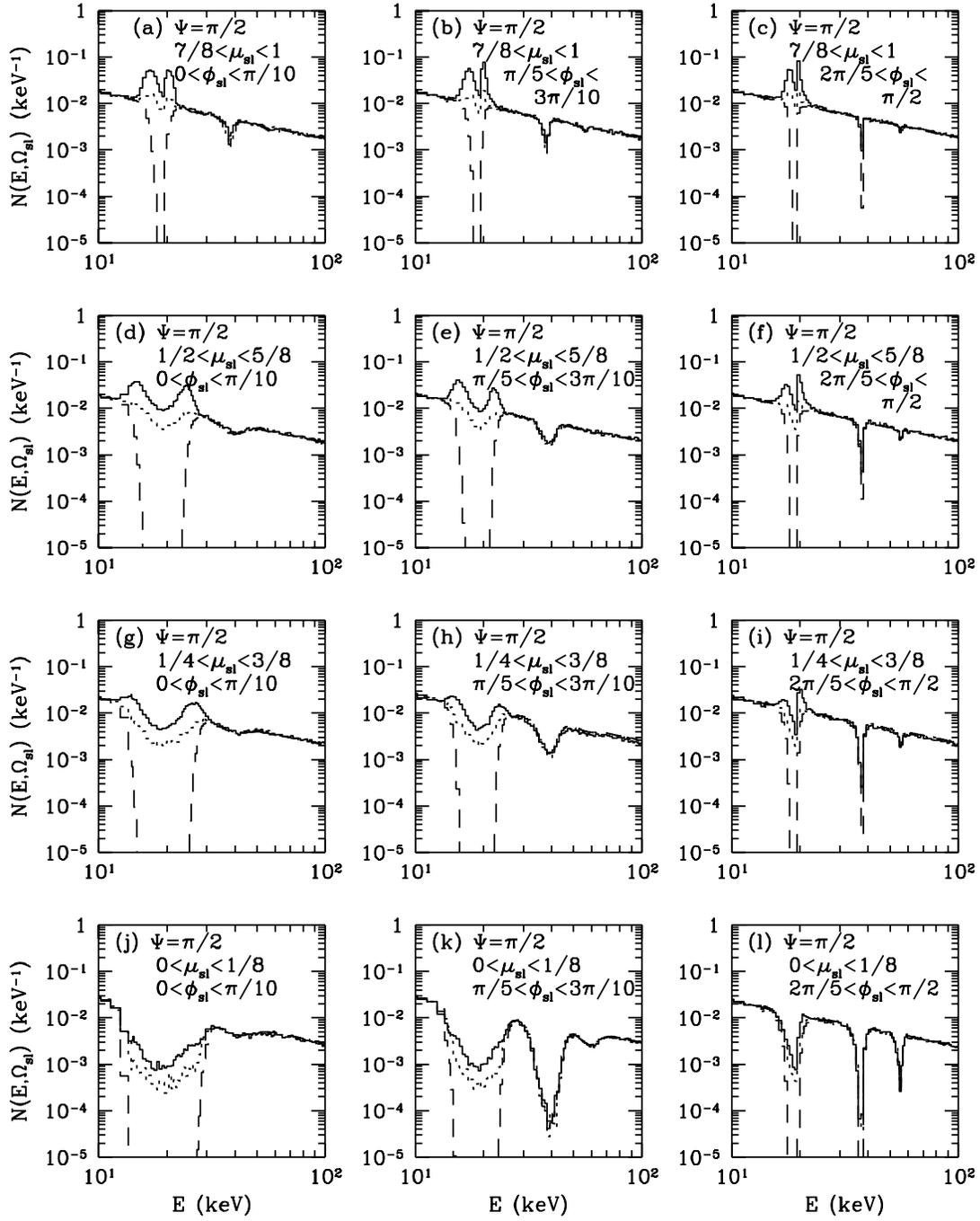
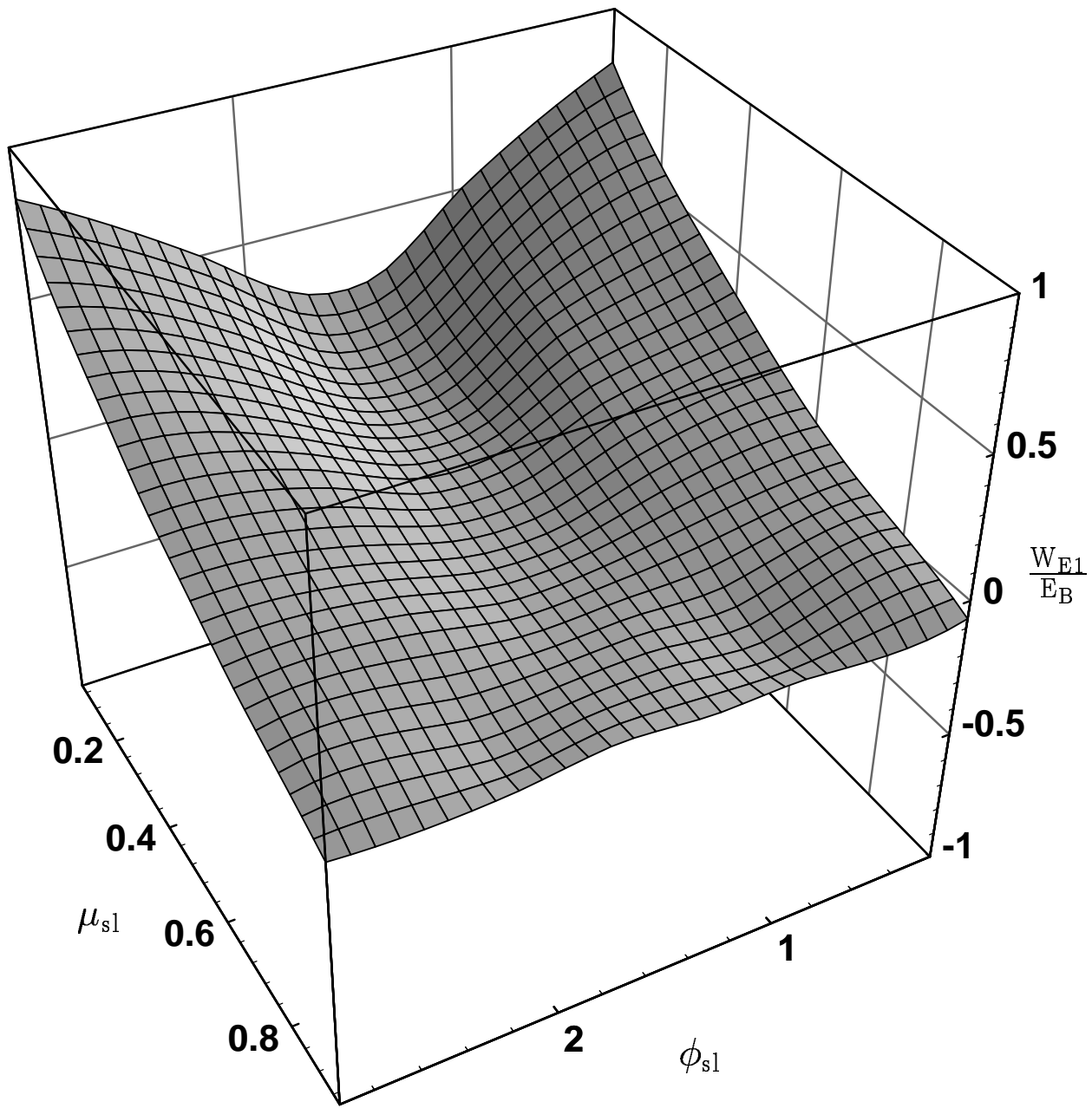
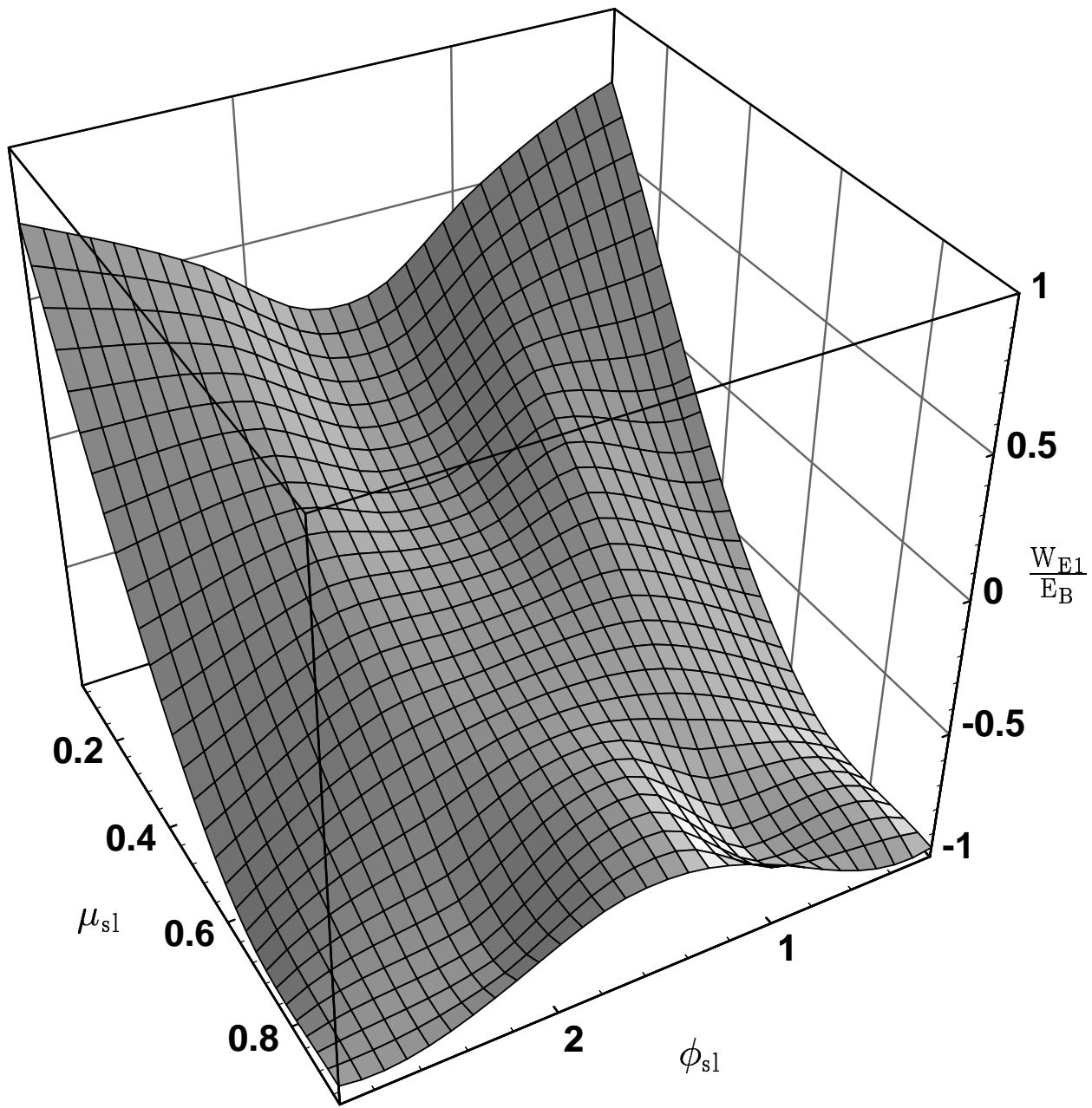
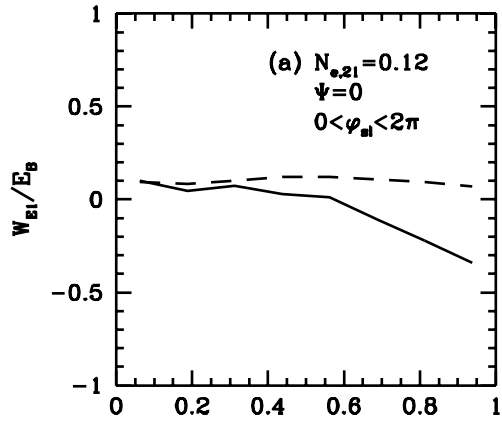
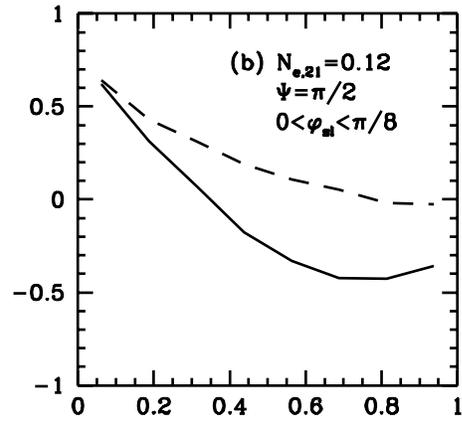
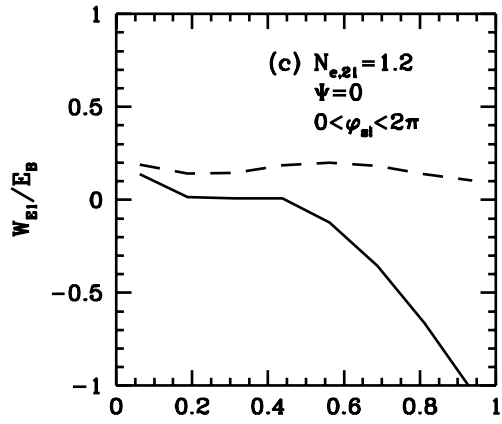
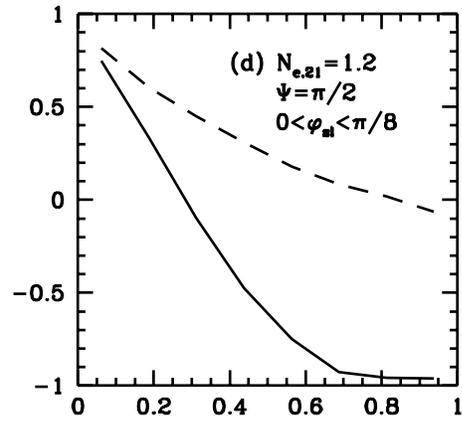
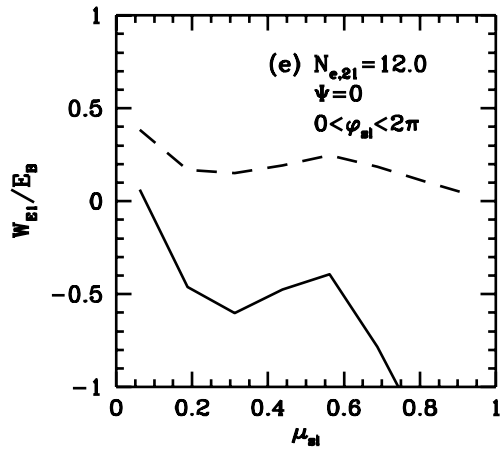
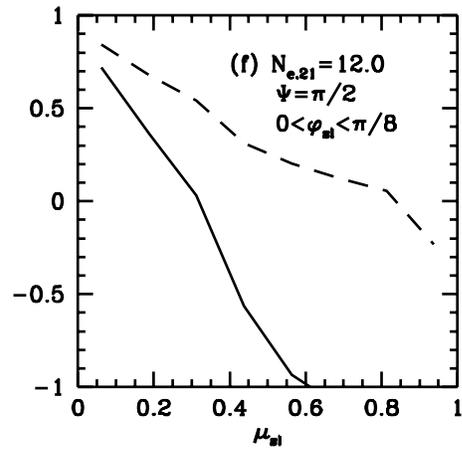
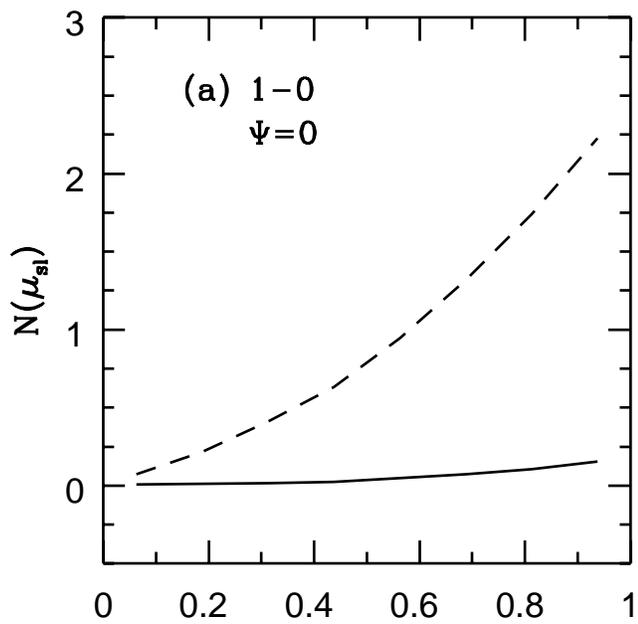
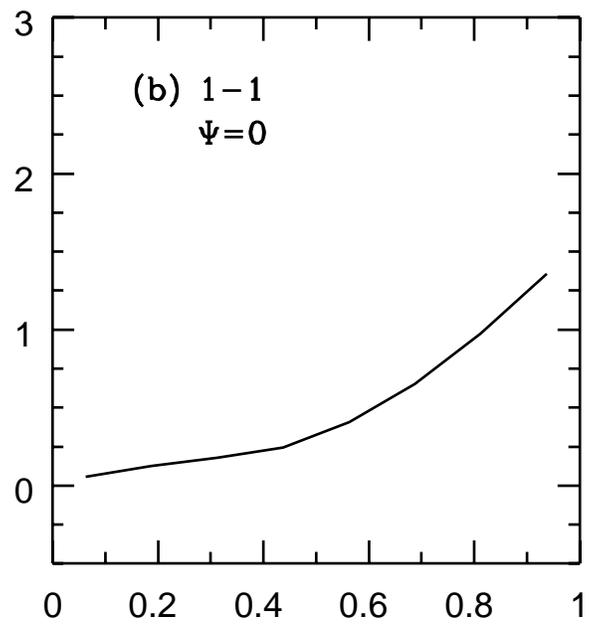
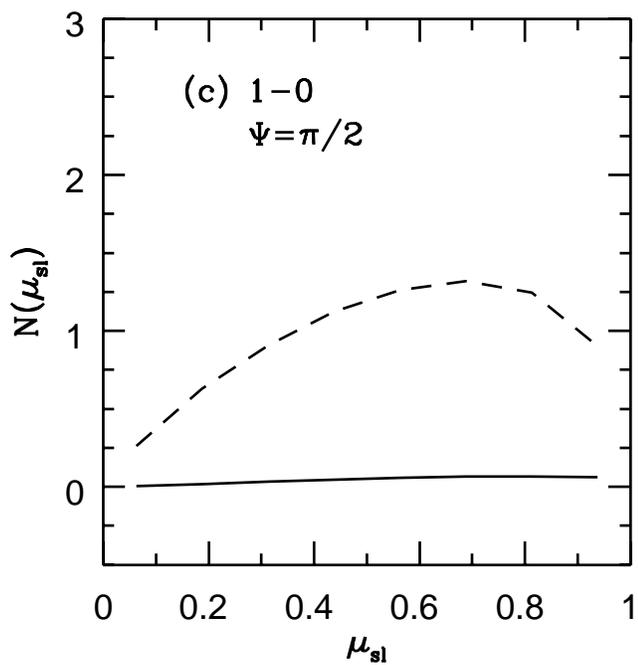
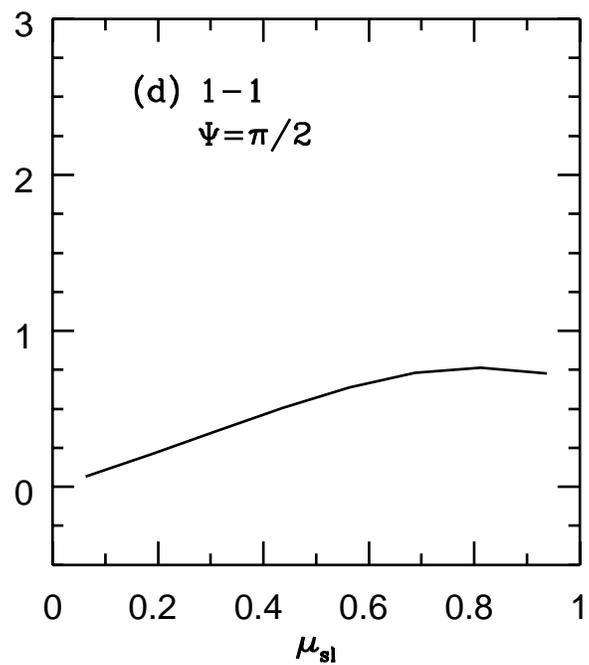
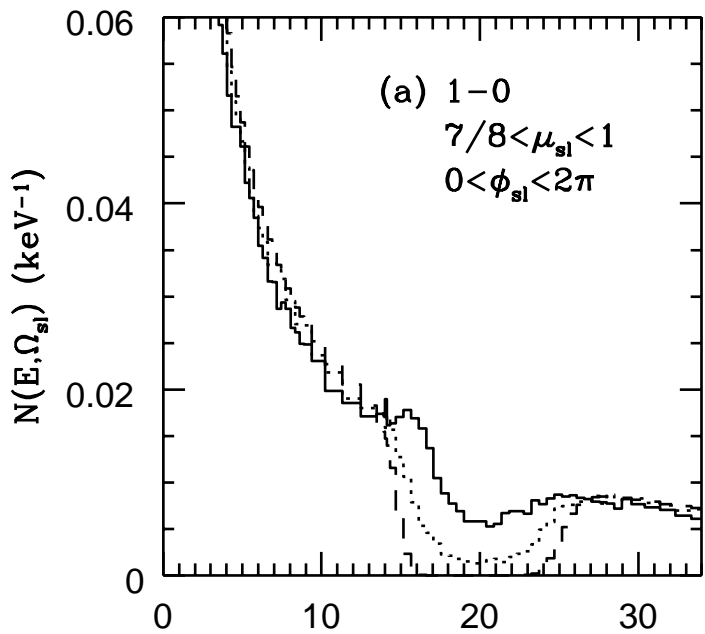
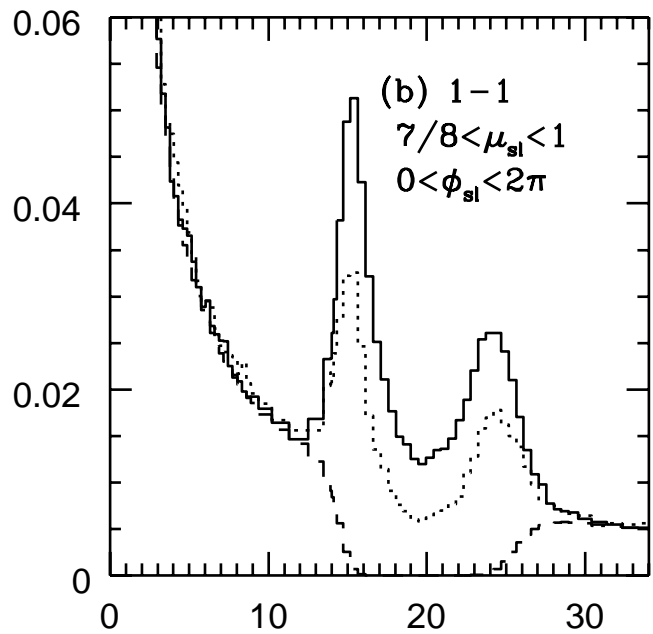
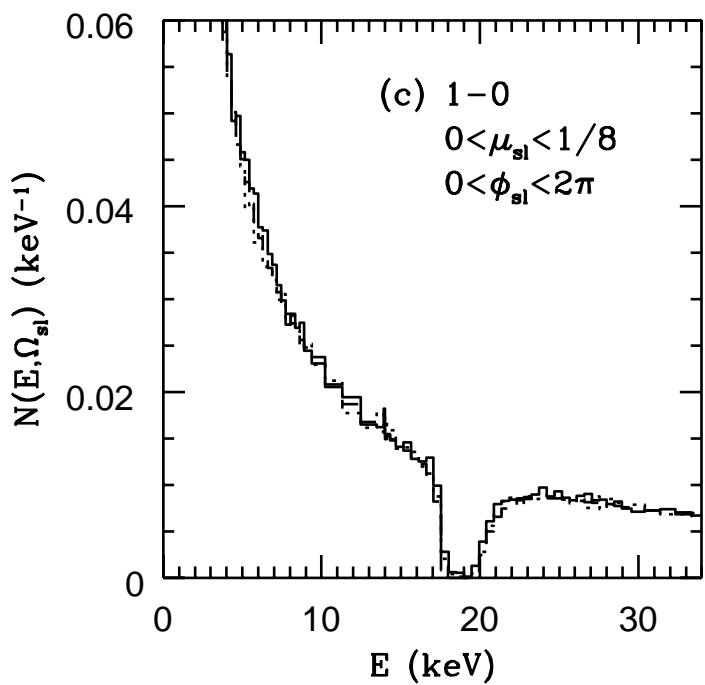
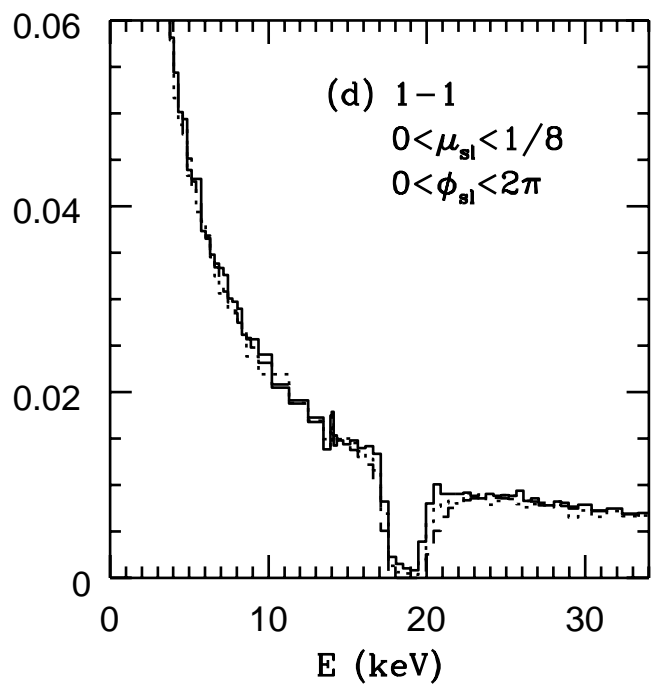
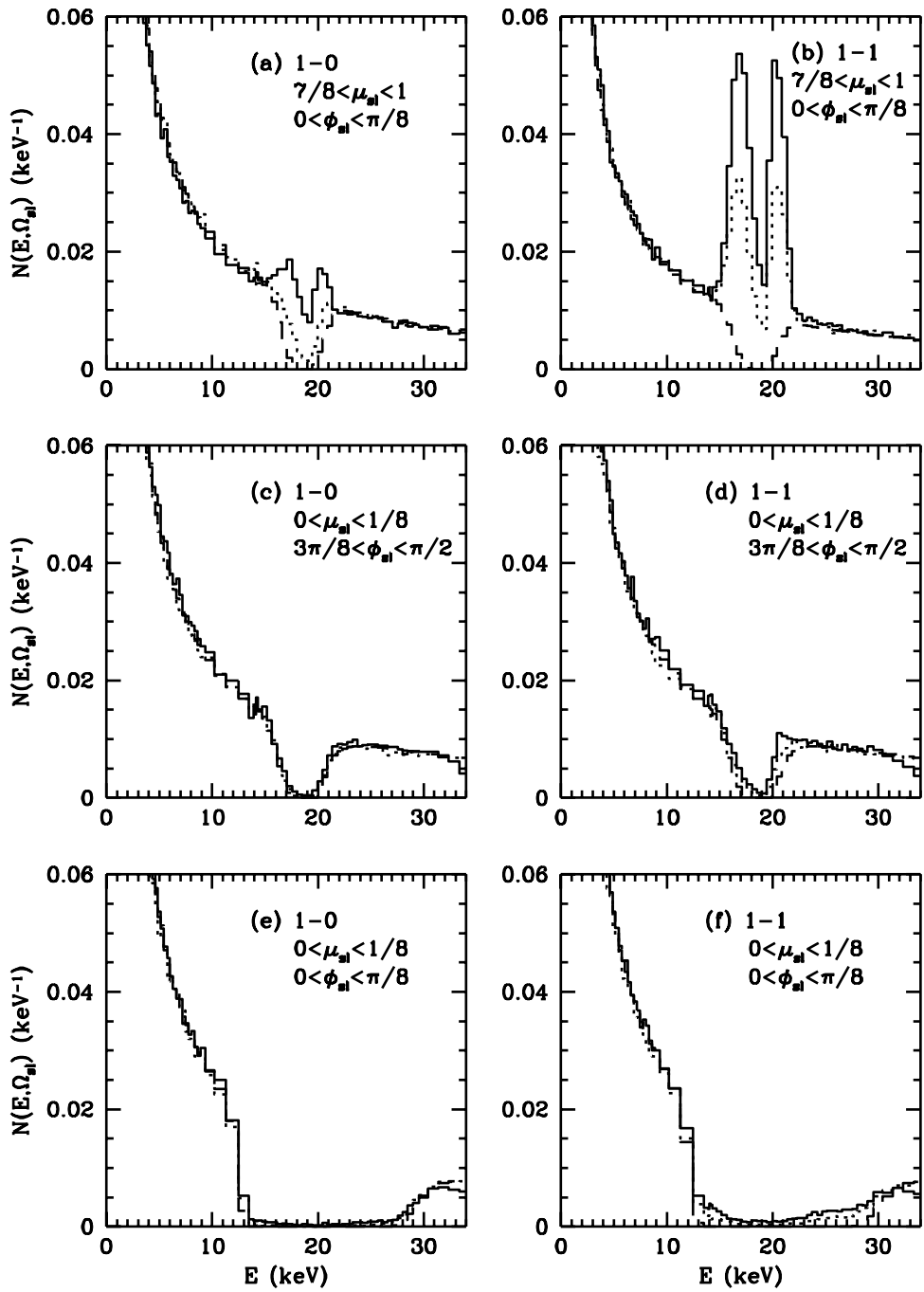
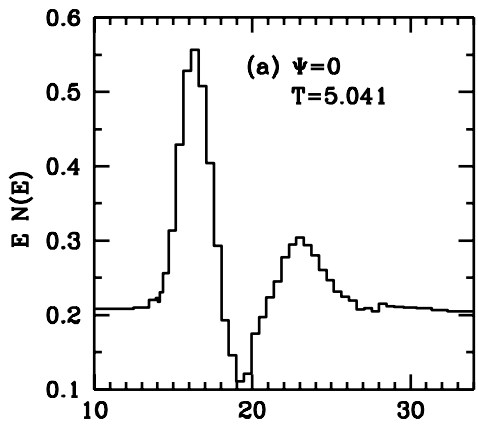
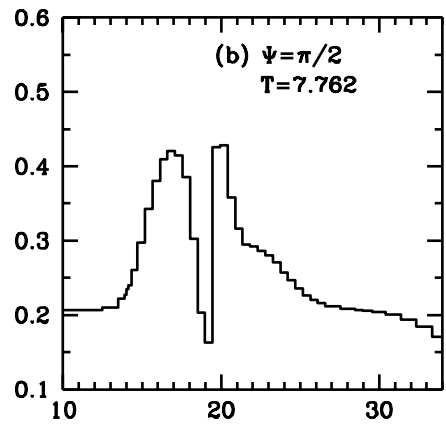
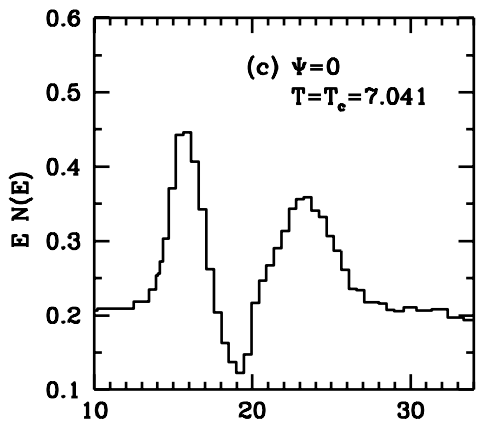
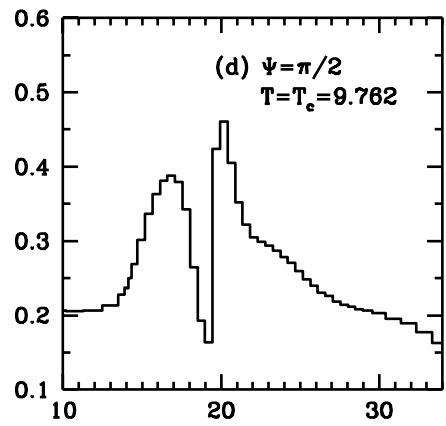
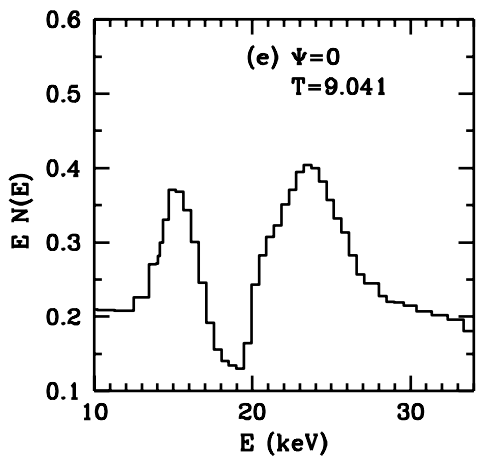
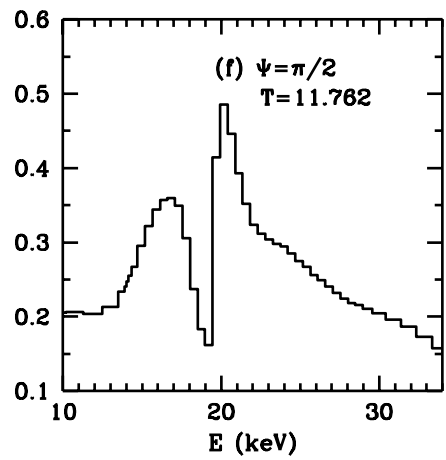
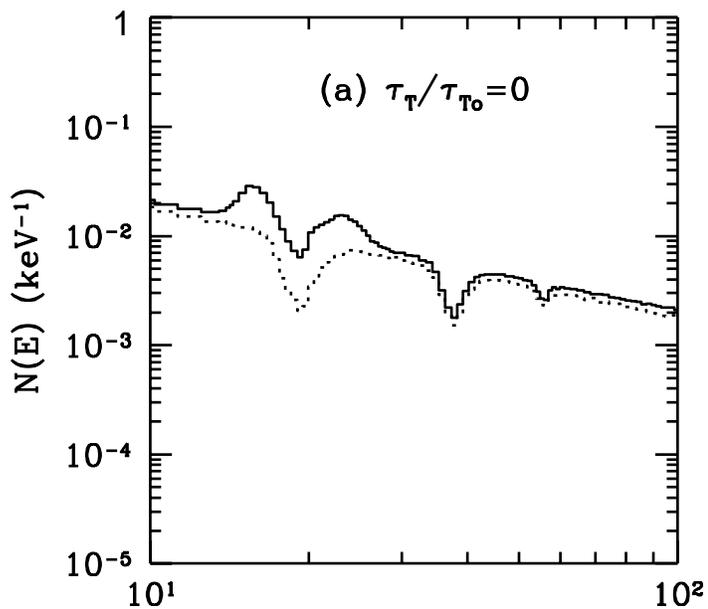
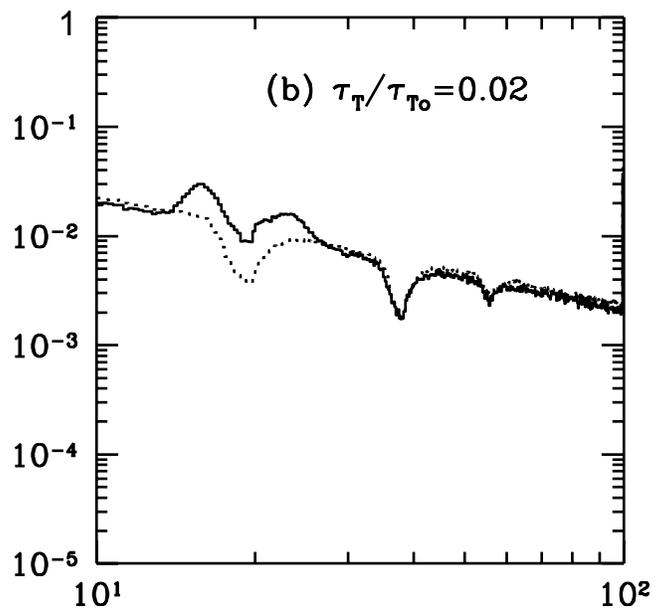
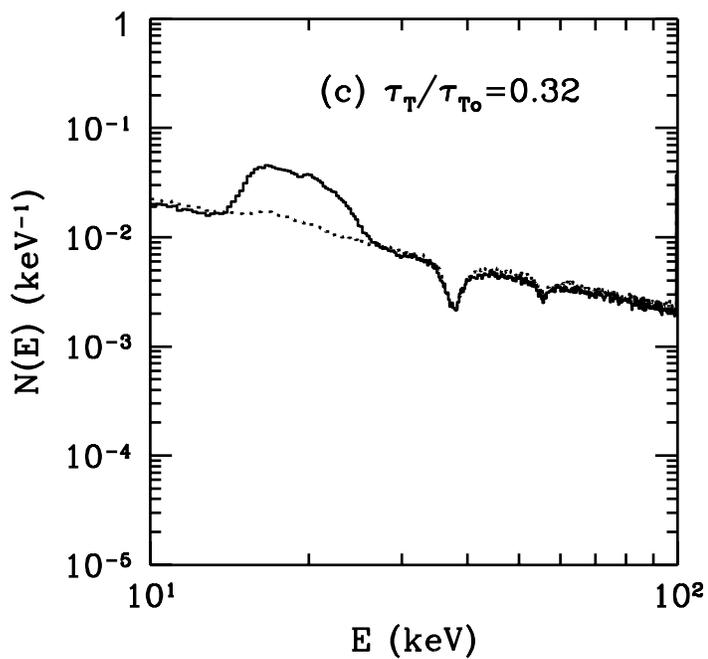
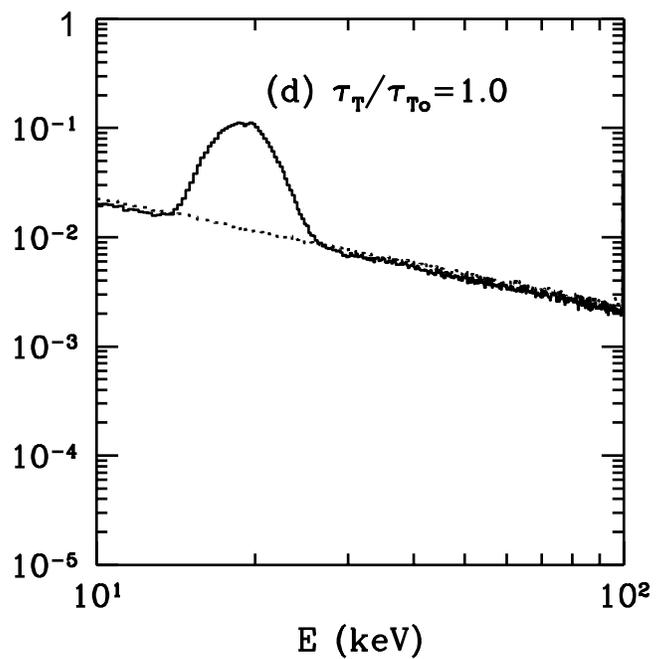
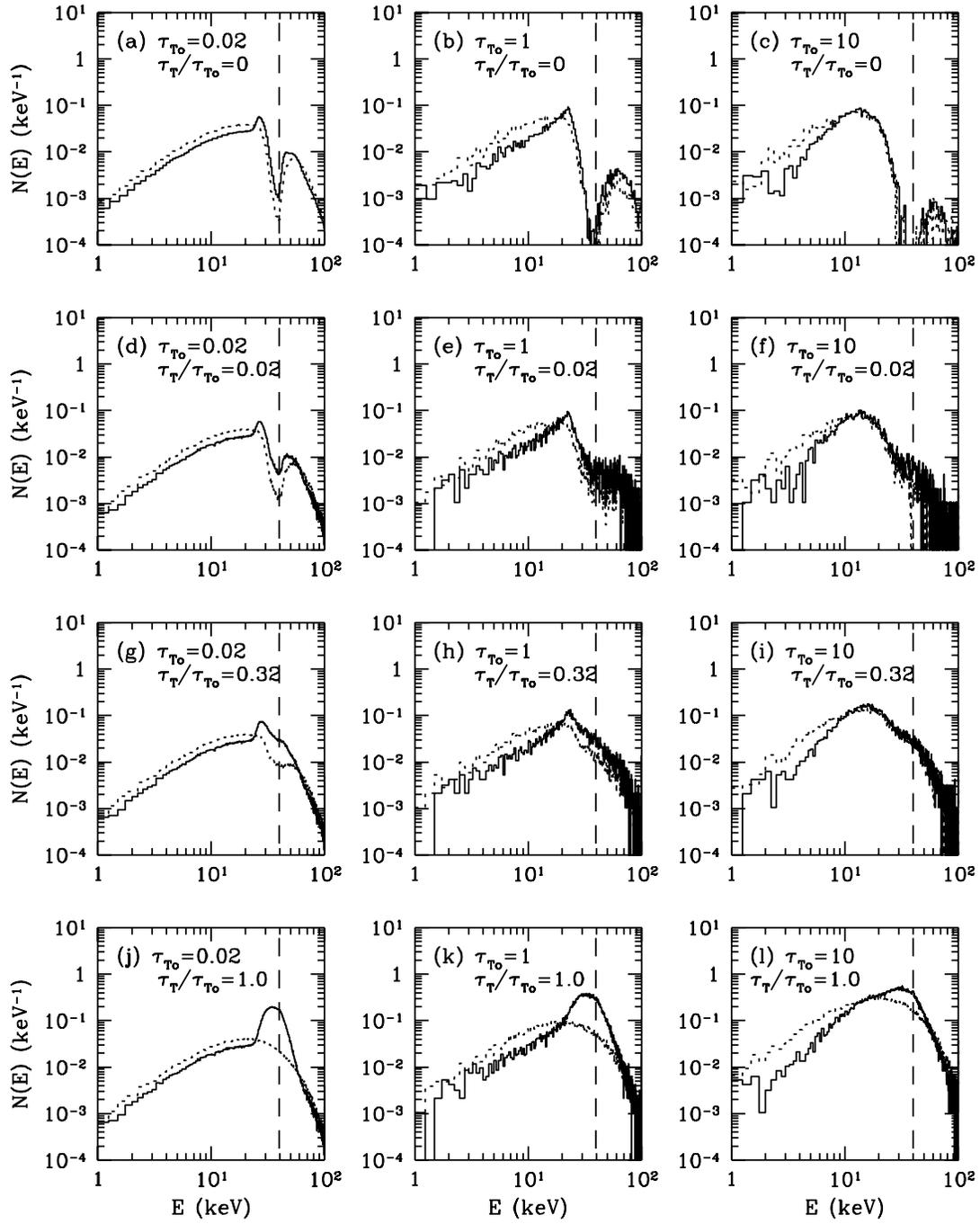
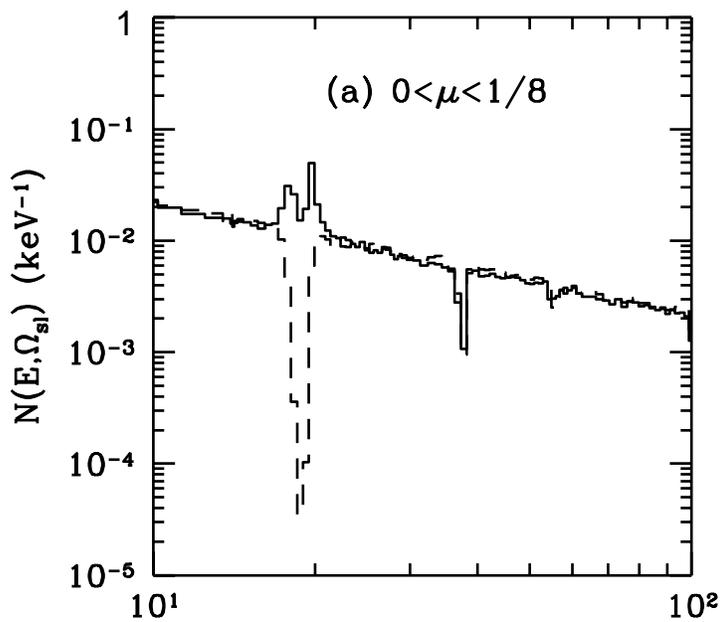
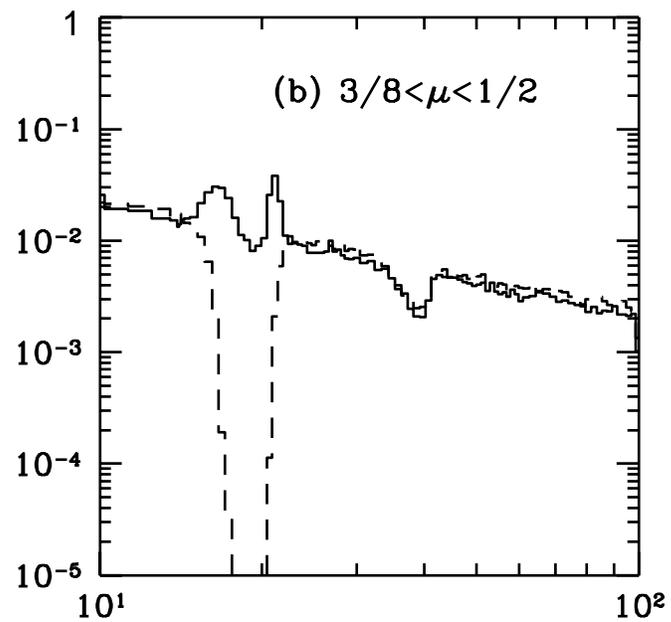
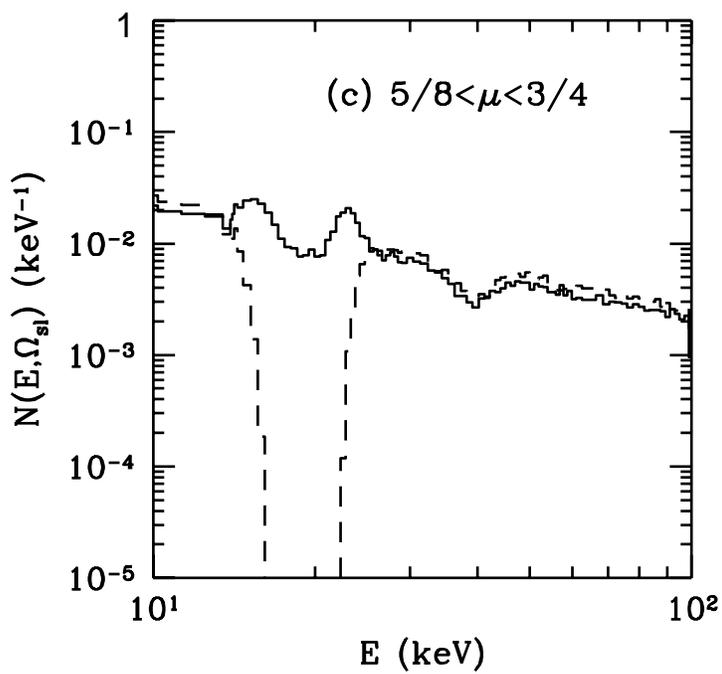
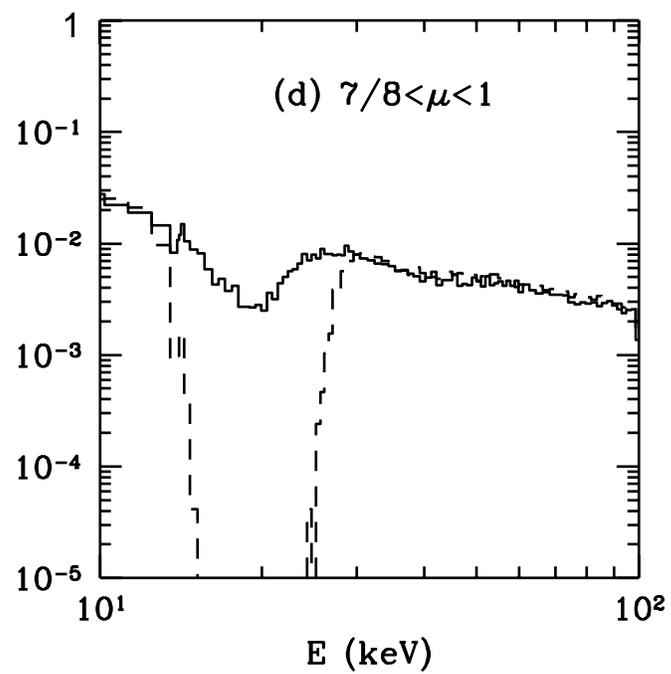
\end{document}